\documentclass[twocolumn,twocolappendix]{aastex631}
\usepackage{{booktabs}}
\usepackage{rotating}
\usepackage{longtable}

\newcommand{\gwsn}{\texttt{GWSkyNet}}

\newcommand{\gwsnm}{\texttt{GWSkyNet-Multi}}
\newcommand{\gwsnmm}{\texttt{GWSkyNet-Multi~II}}

\begin{document}

\title{GWSkyNet-Multi II: an updated machine learning model for rapid classification of \\ gravitational-wave events}

\correspondingauthor{Nayyer Raza}
\email{nayyer.raza@mail.mcgill.ca}

\author[0000-0002-8549-9124]{Nayyer Raza}
\affiliation{Department of Physics, McGill University, 
3600 rue University, Montr\'{e}al, QC H3A2T8, Canada}
\affiliation{Trottier Space Institute at McGill, 
3550 rue University, Montr\'{e}al, QC H3A2A7, Canada}

\author[0009-0009-6826-4559]{Man Leong Chan}
\affiliation{Department of Physics and Astronomy, University of British Columbia, 
Vancouver, BC V6T1Z1, Canada}

\author[0000-0001-6803-2138]{Daryl Haggard}
\affiliation{Department of Physics, McGill University, 
3600 rue University, Montr\'{e}al, QC H3A2T8, Canada}
\affiliation{Trottier Space Institute at McGill, 
3550 rue University, Montr\'{e}al, QC H3A2A7, Canada}

\author[0000-0003-2242-0244]{Ashish Mahabal}
\affiliation{Division of Physics, Mathematics and Astronomy, California Institute of Technology, Pasadena, CA 91125, USA}
\affiliation{Center for Data Driven Discovery, California Institute of Technology, Pasadena, CA 91125, USA}

\author[0000-0003-0316-1355]{Jess McIver}
\affiliation{Department of Physics and Astronomy, University of British Columbia, Vancouver, BC V6T1Z1, Canada}

\author[0000-0001-9290-8603]{Audrey Durand}
\affiliation{Department of Computer Science and Software Engineering, Universit\'{e} Laval, Qu\'{e}bec, QC G1V0A6, Canada}
\affiliation{Canada CIFAR AI Chair, Mila, Montr\'{e}al, QC H2S3H1, Canada}

\author[0009-0002-0606-7516]{Alexandre Larouche}
\affiliation{Department of Computer Science and Software Engineering, Universit\'{e} Laval, Qu\'{e}bec, QC G1V0A6, Canada}

\author[0000-0003-3832-8856]{Hadi Moazen}
\affiliation{Department of Computer Science and Software Engineering, Universit\'{e} Laval, Qu\'{e}bec, QC G1V0A6, Canada}

\shorttitle{\gwsnm ~II}
\shortauthors{Raza et al.}
 
\begin{abstract}

Multi-messenger observations of gravitational waves and electromagnetic emission from compact object mergers offer unique insights into the structure of neutron stars, the formation of heavy elements, and the expansion rate of the Universe. With the LIGO-Virgo-KAGRA (LVK) gravitational-wave detectors currently in their fourth observing run (O4), it is an exciting time for detecting these mergers. However, assessing whether to follow up a candidate gravitational-wave event given limited telescope time and resources is challenging; the candidate can be a false alert due to detector glitches, or may not have any detectable electromagnetic counterpart even if it is real. \gwsnm ~is a machine learning model developed to facilitate follow-up decisions by providing real-time classification of candidate events, using localization information released in LVK rapid public alerts. Here we introduce \gwsnmm, an updated model targeted towards providing more robust and informative predictions during O4 and beyond. Specifically, the model now provides normalized probability scores and associated uncertainties for each of the four corresponding source categories released by the LVK: glitch, binary black hole, neutron star-black hole, and binary neutron star. Informed by explainability studies of the original model, the updated model architecture is also significantly simplified, including replacing input images with intuitive summary values that are more interpretable. For significant event alerts issued during O4a and O4b, \gwsnmm ~produces a prediction that is consistent with the updated LVK classification for 93\% of events. The updated model can be used by the community to help make time-critical follow-up decisions.

\end{abstract}


\section{Introduction} \label{sec:intro}

The LIGO-Virgo-KAGRA (LVK) gravitational-wave (GW) observatories \citep{Acernese2015,Aasi2015,Akutsu2021} have reported 90 significant events involving the merger of binary black holes (BBH), neutron stars and black holes (NSBH), and binary neutron stars (BNS) in published event catalogs across their first three observing runs \citep{Abbott2019_gwtc1,Abbott2023_gwtc3,Abbott2024_gwtc2.1}. With the fourth LVK observing run (O4) currently underway, candidate merger events are being detected regularly ($\sim 200$ significant candidates during O4a and O4b - May 2023 to January 2025), the most promising of which are disseminated by the LVK collaboration in rapid open public alerts\footnote{\url{https://gracedb.ligo.org/superevents/public/O4/}}. This enables follow-up telescope observations of the candidate merger events to capture any electromagnetic (EM) counterpart emission from the GW source\footnote{\url{https://emfollow.docs.ligo.org/userguide/}}.

Merger events that involve a neutron star (NS) are prime candidates for having associated EM bright emission, as was observed for the BNS merger GW170817 \citep{Abbott2017_gw170817}, and the subsequent kilonova AT2017gfo \citep{Abbott2017_multimessenger} and short gamma-ray burst GRB 170817A \citep{Abbott2017_grb170817a}. Multi-messenger events are scientifically rich and offer vast discovery potential, giving insights into varied phenomena such as heavy element nucleosynthesis, neutron star structure, and the expansion rate of the universe \citep{Abbott2017_H0} (see, e.g., \cite{Branchesi2021} for a review of EM counterparts to GW sources). With a single event offering such a wealth of information, other follow-up detections will enable more robust scientific investigations of these types of mergers.

However, telescope time and resources for follow-up are limited, and require careful consideration of the candidate event. Terrestrial sources and chance noise fluctuations can mimic real astrophysical merger events in the LVK detectors \citep{Abbott2018, Abbott2020}, impact the source parameter estimation when in the vicinity of astrophysical signals \citep[e.g.,][]{DalCanton2014, Pankow2018, Powell2018, Hourihane2022, Macas2022, Payne2022, Ghonge2024, Udall2024}, and lead to the release of false alarms. Indeed, in the third observing O3, there were 77 rapid public alerts issued for compact binary coalescence (CBC) events\footnote{\url{https://gracedb.ligo.org/superevents/public/O3/}}, but subsequent detailed analysis determined only 43 of these (56\%) were confidently astrophysical in nature (i.e., included in the GWTC-2.1 and GWTC-3 catalogs). Follow-up observations of false alerts can waste precious telescope resources and time.

Even for the astrophysical events, the potential for having a detectable EM counterpart depends on the type of merger. BNS mergers are the most promising, and are known to produce short gamma-ray bursts and kilonovae, as was observed with GW170817. NSBH mergers may produce an electromagnetic signal (kilonova, GRB) if the NS is tidally disrupted by the black hole (BH) before merger, the probability of which is determined by the mass ratio of the two objects, the aligned spin components, and the NS equation of state \citep[e.g.,][]{Barbieri2020}. To-date no EM counterpart to an NSBH merger has been detected (e.g., \cite{Vieira2020}). On the other hand, BBH mergers in the stellar mass range are not typically expected to produce any counterpart, due to the absence of baryonic matter in the merger surrounding \citep[see, e.g.,][]{Branchesi2021}. For mergers involving neutron stars, the EM counterpart, in particular the kilonova, is also expected to be a transient phenomena that can brighten and fade on the order of hours to weeks across ultraviolet to radio wavelengths, and so the follow-up observations are also time sensitive \citep[see, e.g., the review of kilonovae in][]{Metzger2019}. A fast automated classifier of the candidate LVK events that is open to the public can thus serve as a powerful complementary tool to help make time-critical follow-up decisions.

\gwsn ~\citep{Cabero2020} is a convolutional neural network (CNN) based glitch-vs-real classifier developed for O3 events, which has been further refined and updated by \cite{Chan2024} for O4 events and implemented in the LVK low-latency alert pipeline. The glitch-vs-real model was expanded upon in \gwsnm ~\citep{Abbott2022} as a series of three one-vs-all classifiers to further classify the real (astrophysical) sources as BBH or mergers involving NS (BNS+NSBH). In all iterations the models use low-latency sky map information and associated metadata generated by the rapid localization pipeline \texttt{BAYESTAR} \citep{Singer2016}, with the CNN implemented to extract its own features from the localization images. The classifiers have been shown to perform well on O3 alerts, and are currently being used to make predictions for O4 event alerts\footnote{\url{https://nayyer-raza.github.io/projects/GWSkyNet-Multi/}} \citep{Chan2024}.

In this work we update the \gwsnm ~classifier for O4 events, motivated by the findings in \cite{Raza2024} which explained the model's predictions and identified its limitations and biases. In particular, this included the learned model bias for associating events involving the Virgo detector with astrophysical sources (as compared to glitches), having a discrepancy for annotating the input detector network (online vs triggered), and being insensitive to the signal-to-noise factor of the event. We aim to modify or remove the inputs to the model that do not contribute to the predictions, and wholly update the training dataset to include a more representative sample of LVK events. Additionally, we significantly simplify the model architecture, while making the predictions more informative and nuanced by providing normalized probability scores and uncertainties for the four classes: glitch, BBH, NSBH, and BNS. These updates make predictions for events occurring in LVK O4 and beyond more accurate and robust to the type of event.

The organization of the paper is as follows. In Section \ref{sec:methods} we describe the updates made to the model architecture, the training data, and the general model training procedure. In Section \ref{sec:results} we evaluate the model's performance on test set data and O3 alerts, analyzing the misclassified and high interest events. Predictions for events in the current O4 run are presented in Section \ref{sec:O4_predictions}, along with recommendations for using the model. In the final section we summarize and offer concluding remarks.

\section{Model updates} \label{sec:methods}

\subsection{Updates to architecture} \label{sec:arch_updates}

We make significant changes to the architecture of \gwsnm ~in our effort to make the model simpler and the inputs more interpretable while still maintaining the model's performance. Motivated by our findings in \cite{Raza2024}, in the current study we experimented with modifying each input and each branch of the \gwsnm ~model to study its impact on the model's performance, and to find the best representations of the data that made the classifications possible. Here we describe the changes made to arrive at the final model, which we call \gwsnmm.

For the model inputs we remove the sky localization images and the 3D volume projection images (along with their pixel normalization inputs). The original \gwsnm~models were developed with these images given as inputs with the expectation that the convolutional branches in the model would extract useful features from them. In updating the model with the aim to make the inputs more interpretable, we replace each of the images with a representative numerical value that is more physically intuitive. For the sky map image this corresponds to the 90\% credible interval of the sky localization area (in $\mathrm{deg^2}$), since we found in \cite{Raza2024} that the model was focusing on the size of the localization region in the sky map images to distinguish between astrophysical and glitch events. The sky area represents how well localized the source is on the sky for follow-up observations with telescopes. For the volume projection images we use the 90\% credible interval of the 3D volume localization (in $\mathrm{Mpc^3}$), as this value is related to the sky localization area but also takes into account the distance, and thus indicates the survey volume that would need to be investigated in follow-up observations. The increased input interpretability comes not only from the fact that the model needs to do less work to understand the summary values instead of the images, but also from the fact that these summary values are more familiar and intuitive to the human user. Adding values for other credible interval percentages (for example the 50\% localization area or volume) does not increase the performance of the models. 

For the distance inputs we replace the maximum distance with the standard deviation of the distance (which was previously used to calculate the maximum distance), as a direct measure of the localization uncertainty. Thus the volume (3D), sky area (2D), and distance uncertainty (1D) values offer different representations of the common underlying localization that \texttt{BAYESTAR} computes.

Finally, we re-normalize the Bayes signal vs noise ratio (Log BSN) so that it is clipped at a maximum value of 100; events with Log BSN values of $> 100$ are thus replaced with the value Log BSN $= 100$. While we found in \cite{Raza2024} that the Log BSN factor was not contributing to the model outputs, after clipping the extreme values we find that it can be a useful discriminant between astrophysical events and glitches.

\begin{figure*}
\includegraphics[width=\textwidth]{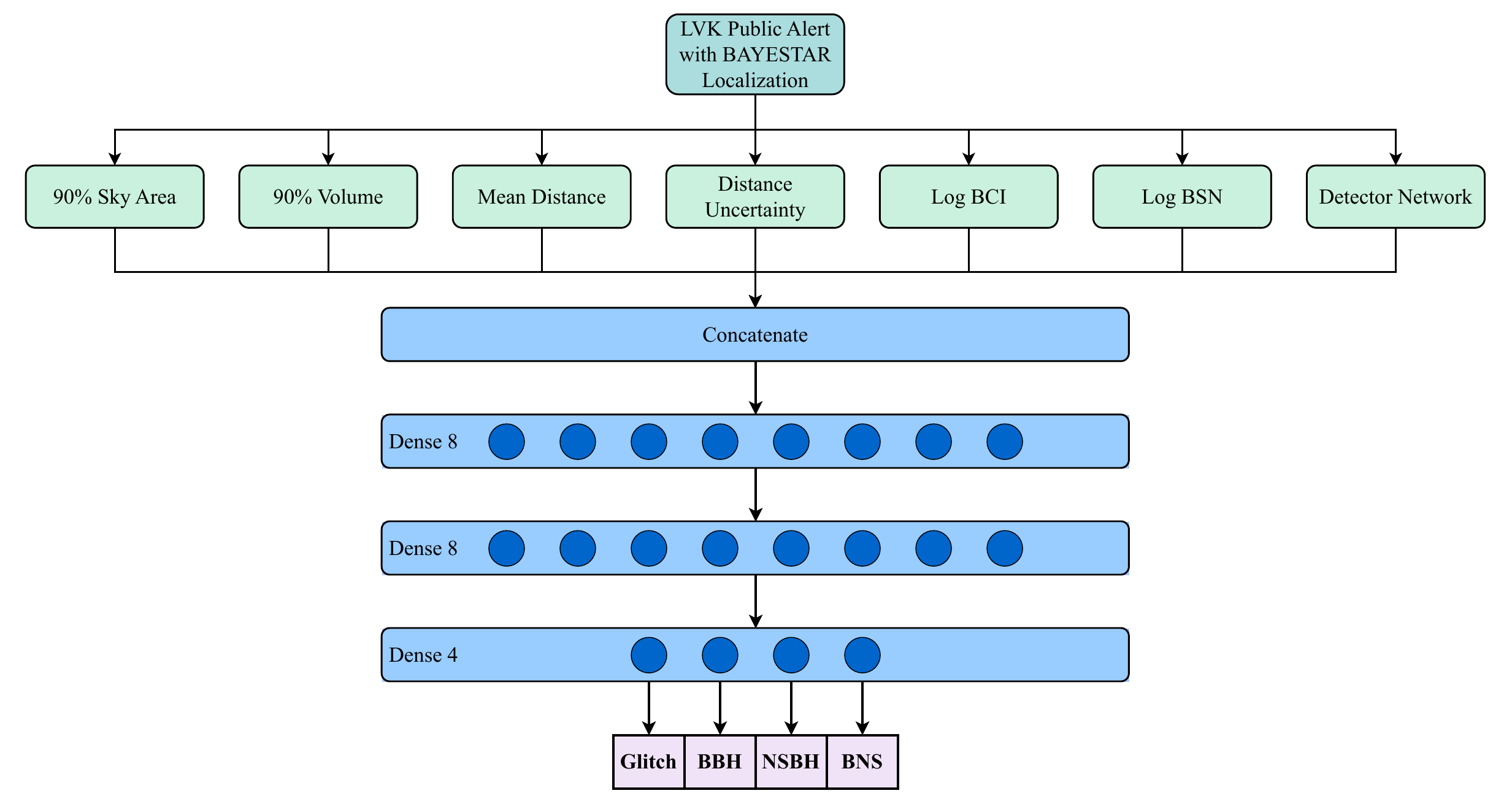}
\caption{The updated model architecture for \gwsnmm, showing the flow of information from the LVK public alert to the final classification by the model. Each input corresponds to a single numerical value, except for the detector network which is a vector of 3 binary (0 or 1) values, representing the observing state of the three detectors. The total 9 inputs are concatenated and fully connected to subsequent dense layers. The dense layers use a \textit{ReLU} (rectified linear unit) activation function, while the final layer uses a \textit{softmax} activation function and outputs four classification probabilities, corresponding to the source being a glitch, BBH, NSBH, or BNS. The total number of trainable parameters in the updated model is 188, compared to a total of 11058 parameters in \gwsnm, representing a factor of $\sim 60$ reduction in the complexity of the model. This results in a significantly shorter training time and a model that is easier to study with explainability tools.
\label{fig:fig1_architecture}}
\end{figure*}

With a sizable reduction in the complexity of the input parameters, we are also able to significantly reduce the complexity of the model architecture. The final \gwsnmm ~model, shown in Figure ~\ref{fig:fig1_architecture}, is a single multi-class classifier (as opposed to three one-vs-all classifiers in \gwsnm) and has only three layers following the concatenated inputs: 1) a dense layer of 8 neurons, 2) a second dense layer of 8 neurons, and 3) the final output dense layer of 4 neurons, each of which outputs the probability of the event belonging to one of the four classes: Glitch, BBH, NSBH, and BNS. The total number of trainable parameters in the updated multi-class model architecture is 188, compared to a total of 11058 parameters across the 3 one-vs-all models in \gwsnm. This represents a factor of $\sim 60$ reduction in the complexity of the model, without compromising the model performance. The change not only allows a significantly shorter training time (allowing us to explore a wider hyper-parameter space for the same compute time), but also results in a model that is potentially easier to study with explainability tools that can give us insights into the features learned.

We emphasize that the updated model now annotates an additional class label, arising from splitting the NS branch into NSBH and BNS, and provides more information to the end user. By using a \textit{softmax} activation function in the final layer we also get normalized prediction probabilities for each of the class types, which can then be directly compared to each other. The updated model thus provides a more complete breakdown of classifications, and also allows for a direct comparison to the preliminary LVK alert classifications.

\

\subsection{Updates to dataset} \label{sec:data_updates}

\subsubsection{Glitches}
The training dataset for glitch events has been updated to now only include events from the LVK third observing run O3 (compared to O1+O2 for \gwsnm ~as in \cite{Abbott2022}). These events are selected from the trigger list of significant and sub-threshold events that were identified in the final offline analysis of the O3 data with a false alarm rate $\mathrm{FAR < 2/day}$ \citep{LVK2021_gwtc21_data, LVK2021_gwtc3_data}, as outlined in \cite{Abbott2023_gwtc3}. There are a total of 1971 unique events identified in the final O3 catalog (O3a+O3b) across three different modeled matched-filter search pipelines for CBC events: GSTLAL \citep{Cannon2021}, PyCBC \citep{Davies2020}, and MBTA \citep{Aubin2021}. For events that are identified by more than one pipeline, we follow the same procedure as the LVK low-latency analysis\footnote{\url{https://emfollow.docs.ligo.org/userguide/analysis/superevents.html}} in selecting the preferred pipeline event; 3-detector triggers are given preference over 2-detector triggers, and among those, higher signal-to-noise ratio (SNR) events are given preference over lower SNR events. Of the 1971 unique events in O3, we consider a sub-selection based on the following criteria: (i) the estimated probability of astrophysical origin $p_{\mathrm{astro}} < 0.5$, (ii) the pipeline found a trigger in at least two detectors with single detector SNR $\rho_{det} > 4.0$, and (iii) the combined network SNR of the event $\rho_{net} > 6.5$. A total of 1868 unique events (95\%) pass these criteria and are included in our final dataset as glitch events.

There is a possibility that some of the 1868 events that we annotate as glitches are in fact weak astrophysical signals, and thus there is some contamination in our glitch training data. However, this number is expected to be low: for the sub-threshold candidate events (which make up almost all of our glitch dataset) it is estimated that $\lesssim 1\%$ of events are astrophysical \citep{Abbott2023_gwtc3}. We consider this contamination rate to be low enough as to not have any significant impact on the model training or how it might learn to differentiate between glitches and astrophysical events. As the LVK detectors undergo upgrades between each run, the sensitivity of the detectors and their noise profile naturally also change, which means that we do not fully capture all of the new kinds of transient glitch events that could occur in O4. Nevertheless, by using the latest publicly available datasets, the 1868 glitches from O3 in our updated training data are more representative of the instrumental artifacts expected in O4, as compared to using glitches from O1 and O2.

\subsubsection{Astrophysical events}
The astrophysical events (BBH, NSBH, BNS) are simulated events that are injected into Gaussian noise colored by the power spectral densities (PSDs) of the detectors during O3. We generate 1868 simulated events for each of the three astrophysical event types, with the number of events chosen to match the number of glitch events in the training dataset. The simulated events in the updated dataset largely reflect updates to population model fit parameters compared to those available when \gwsnm ~was first developed \citep{Abbott2022}. The BH masses are sampled from a power law + peak model with parameters determined from the population in GWTC-3 \citep{Abbott2023_gwtc3_population}. For NS masses we follow the distribution determined in \cite{Landry2021} from the population of GW merger events with NS, and sample the masses uniformly in the range $[1~M_{\odot}, 2.5~M_{\odot}]$ with random pairing (the maximum NS mass of $2.5~M_{\odot}$ is chosen as a compromise between the two different values determined in \cite{Landry2021}, depending on the exclusion or inclusion of the event GW190814 in the model fits: $m_{max} = 2.0^{+0.4}_{-0.3}$ and $m_{max} = 2.7^{+0.2}_{-0.2}$). For BBH events we constrain the mass ratio according to the power law + peak distribution in \cite{Abbott2023_gwtc3_population}, and for NSBH events the mass ratio is constrained to be $q=m_2/m_1 \geq 1/30$, determined by the 99\% confidence interval upper limit of the mass ratio for NSBH events from binary evolution simulations in \cite{Drozda2022}. For the spins we assume a simple uniform distribution with aligned spins in the range [0, 0.99] for BHs and [0, 0.05] for NSs. Table~\ref{table:astrophysical_population_params} summarizes the parameter distributions that we sample from to generate the simulated waveforms for each merger type. The distances for events are sampled uniformly in volume, and the remaining extrinsic source parameter angles (longitude, latitude, inclination, polarization) are sampled uniformly.

\begin{table*}
\centering
\vspace*{0.2cm}
\hspace*{-1.9cm}
\begin{tabular}{ |c|c|c|c| }
\toprule
\toprule
\textbf{Merger Type} & $\mathbf{Mass ~[M_{\odot}]}$ & \textbf{Spin (aligned)} & \textbf{Mass Ratio}\\
\midrule
\midrule
BBH & BH: Power+Peak(5.1, 88) & BH: Uniform(0, 0.99) & $\beta_q = 1.1$ \\
    & BH: Power+Peak(5.1, 88) & BH: Uniform(0, 0.99) & \\
\hline
NSBH & BH: Power+Peak(5.1, 88) & BH: Uniform(0, 0.99) & $q \geq 1/30$ \\
     & NS: Uniform(1.0, 2.5) & NS: Uniform(0, 0.05) &  \\
\hline
BNS & NS: Uniform(1.0, 2.5) & NS: Uniform(0, 0.05) & random pairing \\
    & NS: Uniform(1.0, 2.5) & NS: Uniform(0, 0.05) &  \\
\bottomrule
\end{tabular}
\caption{Summary of the astrophysical source parameter distributions used to generate the simulated waveform signals for BNS, NSBH, and BBH merger events in our training and testing data set. Power+Peak refers to the power law + peak population mass model outlined in \cite{Abbott2023_gwtc3_population}, with $\beta_q$ the spectral index for the power law of the mass ratio distribution. The $q=m_2/m_1 \geq 1/30$ mass ratio constraint for NSBH mergers is determined from \cite{Drozda2022}.}
\label{table:astrophysical_population_params}
\end{table*}

To simulate the astrophysical event waveforms we use the \texttt{TaylorF2} \citep{Mishra2016} waveform model for events with $m_1 + m_2 < 4~M_{\odot}$, and the \texttt{SEOBNRv4-ROM} \citep{Bohe2017} models otherwise, generated using the LALSuite software \citep{lalsuite2018}. Each waveform is then injected into Gaussian noise colored by the actual PSD of the Hanford, Livingston, and Virgo detectors during one of the confident astrophysical events in O3. We randomly sample from $\sim 40$ such distinct O3 event PSDs, as provided in the GWTC-2.1 and GWTC-3 data releases \citep{LVK2021_gwtc21_data, LVK2021_gwtc3_data}. These PSDs are chosen because they represent clear examples of when the detectors were operating with high enough sensitivity to detect astrophysical events, and because they best approximate the LIGO-Virgo detector configurations/sensitivities required to detect typical astrophysical events in O3. The events are distributed such that the number of events simulated in each detector combination (HLV, HL, HV, or LV) is equal.

The waveform is injected into O3-colored Gaussian noise, as opposed to the actual O3 detector noise, to ensure that these events can be unambiguously classified as astrophysical events for model training/testing, without any chance of containing overlapping non-Gaussian glitch features. We then use the \texttt{ligo.skymap}\footnote{\url{https://git.ligo.org/lscsoft/ligo.skymap}} package to simulate running a matched-filter search pipeline to detect the injected signal and producing the corresponding sky localization map. The first step adds Gaussian measurement error to the optimal SNR timeseries (calculated based on the simulated waveform template) and time of arrival of the event across multiple detectors, and determines which injections would be detected based on SNR thresholds and produces corresponding triggers. The second step then inputs these coincident triggers to the \texttt{BAYESTAR} pipeline to produce the GW event localization map, output as a FITS file. Finally, the data in the \texttt{BAYESTAR} map is processed to calculate the relevant model inputs for \gwsnmm ~(as shown in Figure~\ref{fig:fig1_architecture}, e.g., the 90\% sky localization area). With the $1868 \times 3 = 5604$ simulated astrophysical events, the model can learn the distinguishing features of the underlying sources more robustly.

\subsubsection{SNR threshold and final dataset}
Based on biases identified in \cite{Raza2024} for the type of events included in training the model, we also lower the SNR threshold for events, and modify the detector status input to be in line with the LVK low-latency annotation. In \gwsnm ~the threshold for including events was set so that the matched filter SNR in at least two detectors was $\rho_{det} > 4.5$, and the network SNR $\rho_{net} > 7$. Since events with lower SNR can be part of public alerts, we lower this threshold so that in our updated dataset we have events with $\rho_{det} > 4.0$ and $\rho_{net} > 6.5$. In \gwsnm ~the detector status was based on which detectors had a trigger identified by one of the pipelines. For the glitches, the detector network status is updated to reflect whether or not each individual detector (Hanford, Livingston, or Virgo) was in observing mode at the time of the event, and thus contributed to the sky-localization, regardless of whether one of the pipelines found a corresponding trigger in the detector or not. For example, if an event occurred while all three detectors were in observing mode, but there was a trigger found in only the Hanford and Livingston detectors and not Virgo, this would now be annotated as a three-detector HLV event rather than a 2-detector HL event. Similarly, for the astrophysical events in our dataset, the detector network status corresponds to the detector combination in which we simulate injecting waveforms and use to construct the sky localization, regardless of whether the resulting waveform SNR in each of the detectors would be high enough to trigger a detection pipeline. The detectors that contributed to the sky localization (as opposed to the detectors for which a trigger was found) is part of the information that is made publicly available by the LVK in its alerts, and so the updated data set now replicates the actual alert contents for O4 more closely.

\begin{figure*}
\includegraphics[width=\textwidth]{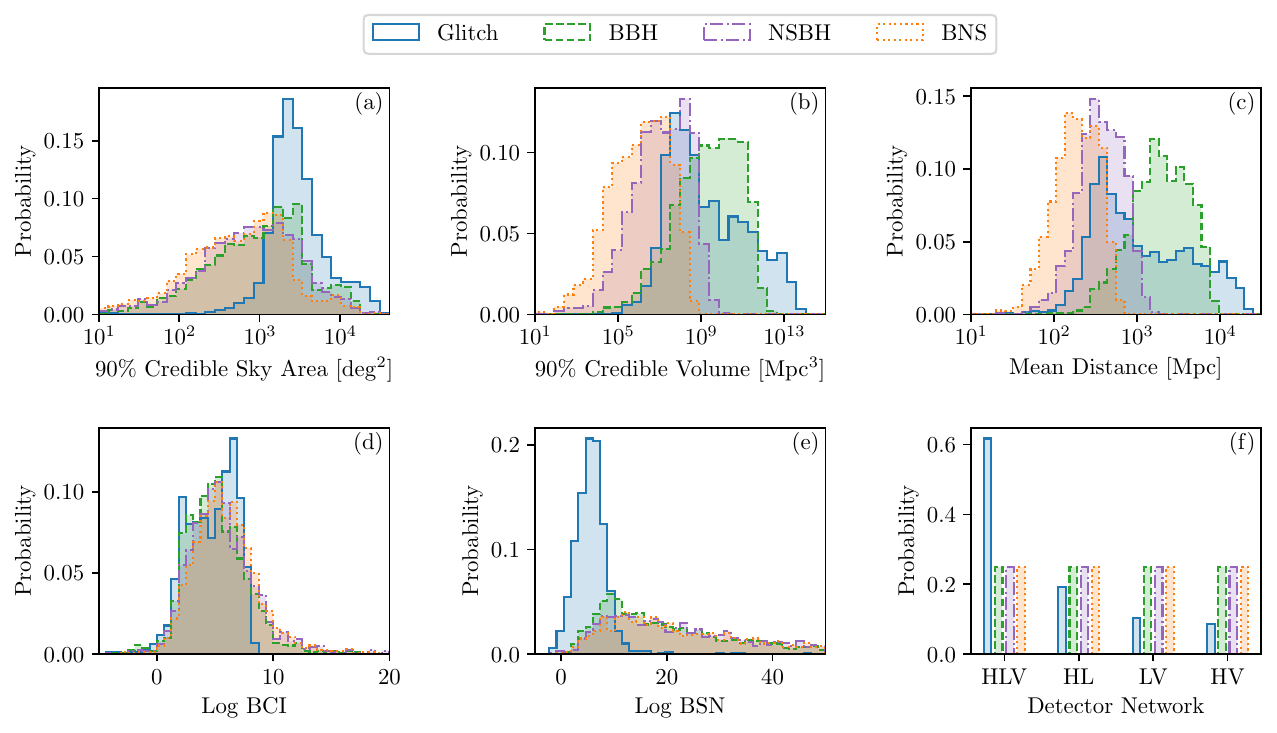}
\caption{Probability distributions of the updated input parameters to \gwsnmm ~over the entire data set of events for the glitch (blue solid line), BBH (green dashed line), NSBH (purple dash-dotted line), and BNS (orange dotted line) sources: (a) sky map localization area, (b) 3D volume localization, (c) estimated mean distance to the source, (d) Bayes factor for coherence versus incoherence, (e) Bayes factor for signal versus noise, and (f) the network of detectors observing during the event. The distributions are distinct for certain input types and thus we expect that the models will learn to use these features to discriminate between source classes. In particular, the sky map localization area and Log BSN factors can potentially distinguish between glitch events and astrophysical sources. While the mean distance estimate and 90\% credible volume could be leveraged to classify between BBH, NSBH, and BNS mergers.
\label{fig:fig2_inputs_distributions}}
\end{figure*}

The resulting distributions of key inputs for the glitch, BBH, NSBH, and BNS events are shown in Figure~\ref{fig:fig2_inputs_distributions}. The glitch events have distinct features compared to the astrophysical events for most inputs, except for the Bayes coherence-vs-incoherence ratio (Log BCI). This is a marked change compared to the Log BCI distributions for the dataset used in \gwsnm~(see Figure 1 in \cite{Raza2024}). The glitches used in the previous dataset had lower Log BCI values, possibly arising from the fact that there was no FAR threshold cutoff used to decide which glitch events to include. Among the astrophysical events, we can see that the mean distance estimate and 90\% credible volume can distinguish between the three merger classes. We thus expect the models to exploit these differences to learn the correct classifications.

As can be seen in Figure~\ref{fig:fig2_inputs_distributions}, some of the input parameters can span several orders of magnitude for the range of events we consider in our dataset. To avoid numerical issues during model training, we pre-process the data by transforming and normalizing the input values: i) for the sky area, volume, and distances we first take the logarithm of the values, and ii) for all inputs we then normalize the values by dividing them by the maximum value in the dataset.

\subsection{Model training and uncertainty estimation} \label{sec:model_training}

We randomly split the 7472 events in our final dataset into three categories: i) 81\% of events are used for training the model, ii) 9\% of events are used for validation during training, and iii) 10\% are a hold-out reserved for testing the model's performance after the training has completed. \gwsnmm ~is developed and trained using \texttt{TensorFlow} \citep{Abadi2015} and \texttt{Keras} \citep{Chollet2015}. We use the \texttt{Adam} optimization algorithm \citep{Kingma2014} with a learning rate of 0.001 and a batch size of 64 to train the model, setting the categorical cross-entropy loss as our objective function to minimize. The maximum number of training epochs allowed is 2000. The performance of the model is evaluated after each training epoch by calculating the cross-entropy loss of the validation set, and the training is stopped early if the validation loss does not decrease for 100 consecutive epochs. This early-stopping based on the validation set performance ensures that the model does not over-fit to the training data. To ensure that the final model is a result of the model learning features from the training set and reaching a global minimum loss, rather than a serendipitous case of initializing with the best weights, we repeat the training five times with different random weight initializations each time. From these five trained models we select the one which has the median accuracy on the validation set (as the representative case, and find only a marginal accuracy difference of $\sim 1-2 \%$ between the highest and median performing models). The training hyperparameter values for the learning rate, batch size, maximum number of epochs, and patience of early-stopping are manually tuned and the final values chosen such that the model's performance on the validation set is optimized (maximum accuracy). During training, we also down-weight the BNS and NSBH events by a factor of 10 and 5, respectively, as we find this helps the models generalize better to events in O3. This is discussed in more detail in Appendix~\ref{sec:appA}.

A single trained model outputs four probability values, but does not give any indication of the uncertainties associated with each probability, which would be a useful measure of the model's confidence in the predictions. In our training we split our dataset so that 90\% of the dataset is used for training and validation, while the remaining 10\% is kept for testing the model after training, with the events randomly shuffled before doing the split. To compute the uncertainty in the model's predictions, we generate 20 randomized train-test splittings of the dataset, and train 20 separate models, one for each randomized split. To compute the predictions for each event, the output probabilities from all 20 models are combined and the probability mean and standard deviation for each class is calculated and reported as the final value. Thus we use an ensemble of 20 models trained on slightly different data to determine the uncertainty on the predictions. When a single classification label is required for an event (for example, to calculate the accuracy of the model) we select the class which has the highest predicted mean probability (a winner-take-all approach).

\subsection{Comparison to alternate models}
Since the updated model inputs are all numerical values (tabular data), we are not constrained to the neural network model architecture (which is known to perform well for image-based inputs), but can also explore alternate supervised learning models that are known to perform well on numerical (tabular) input data. We train two such state-of-the-art models: i) eXtreme Gradient Boosting \citep[XGBoost,][]{Chen2016}, which implements gradient boosted trees, and ii) Explainable Boosting Machine \citep[EBM,][]{Lou2013,Nori2019}, which is a tree-based gradient boosting generalized additive model, and has the advantage of being fully interpretable. We train the XGBoost and EBM models with the same inputs as the neural network, and use \texttt{Optuna} \citep{Akiba2019}, an automatic hyperparameter optimization framework, to select the best model hyperparameters over 5000 trials.

Table~\ref{table:architecture_comparisons} shows the classification performance of XGBoost and EBM, as compared to the neural network, for the final selected models. The results show that while XGBoost and EBM have similar accuracy as the neural network on the test set of events, they perform considerably worse on the O3 events (by $\sim 10-15 \%$). The features learned by the XGBoost and EBM models on the training data thus do not generalize as well to the O3 event alerts, whereas the neural network has learned more generalized features that tranfer well across the two different data sets. We thus proceed with the neural network as our preferred model for \gwsnmm.

\begin{table}
\vspace*{0.2cm}
\centering
\begin{tabular}{|l|c|c|}
\toprule
\toprule
\textbf{Model} & \textbf{Test Accuracy (\%)} & \textbf{O3 Accuracy (\%)} \\
\midrule
\midrule
NeuralNet & $85.1 \pm 1.3$ & 81.3 \\
XGBoost & $85.6 \pm 1.0$ & 66.7 \\
EBM & $82.2 \pm 1.4$ & 72.0 \\
\bottomrule
\end{tabular}
\caption{A comparison of the classification accuracy scores of three different state-of-the-art machine learning models: Neural Network (NeuralNet), eXtreme Gradient Boosting (XGBoost), and Explainable Boosting Machine (EBM). We compare the prediction accuracy for two datasets: the hold-out test set of events which has the same underlying distribution as the training data, and the O3 event alerts which correspond to the events that were released by the LVK during the third observing run (see Section~\ref{sec:results} for more details). The NeuralNet, XGBoost, and EBM models have comparable accuracy for events in the test set, but NeuralNet significantly outperforms XGBoost and EBM for O3 events. We thus select the neural network model as our chosen model and architecture for \gwsnmm.}
\label{table:architecture_comparisons}
\end{table}

\section{Model Validation and Analysis} \label{sec:results}

We evaluate the predictions of the updated model, \gwsnmm, and validate them on two distinct sets of events: i) the 10\% of events from our dataset that were kept as a hold-out test set, and ii) the 75 multi-detector public alerts issued by the LVK during O3.

\subsection{Performance on Test Set Events}

\begin{figure}
\includegraphics[width=\columnwidth]{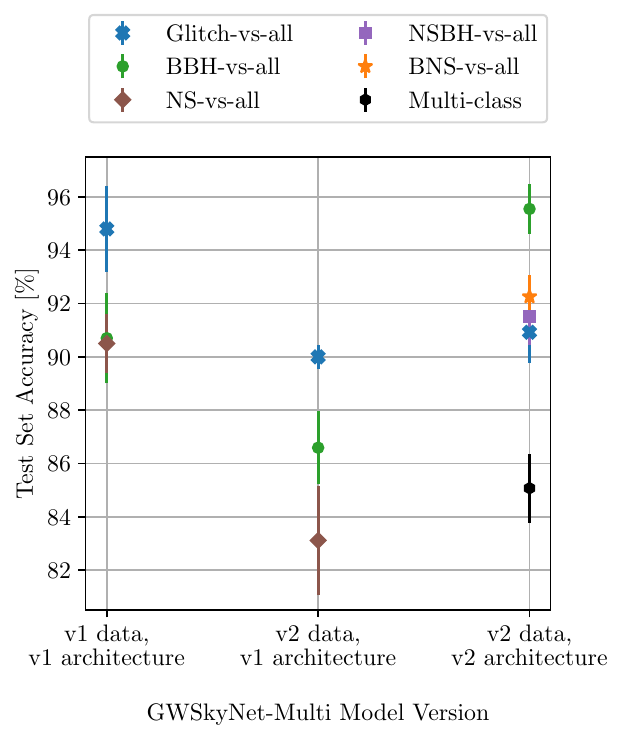}
\caption{Model performance changes as the dataset and architecture are updated, characterized by the test set accuracy. Scores in the left-hand column (v1 data, v1 architecture) are for the original \gwsnm ~model as described in \cite{Abbott2022}. Scores in the right-hand column are for the updated model \gwsnmm ~presented in this work (v2 data, v2 architecture), showing both the full multi-class accuracy and the aggregated one-vs-all accuracy scores for easier comparison. Scores in the center represent the intermediate model which has the same architecture as the original model but has been trained on the updated dataset (v2 data, v1 architecture). Upgrades made to the dataset only lead to generally worse performance for the model, indicating that the v2 dataset is inherently harder to distinguish. Subsequent architecture changes bring the performance back to $\sim 90-95\%$ accuracy when we compare the aggregated one-vs-all models, with a multi-class accuracy of $\sim 85\%$.
\label{fig:fig3_test_accuracy}}
\end{figure}

The accuracy of the multi-class model predictions for the hold-out test set along with a comparison to previous iterations of the model is shown in Figure \ref{fig:fig3_test_accuracy}. For the multi-class model the predicted class label is determined by the class with the highest probability score from the four values. For each of the 20 trained models in our ensemble we only make predictions on the 10\% of events (747 events) that were not in their respective training set to calculate the test accuracy. We then calculate the mean (and standard deviation) of these 20 accuracy scores as our model's overall test-set performance accuracy (and uncertainty on that accuracy). The final \gwsnmm ~model, denoted as the multi-class model in the figure, has an accuracy of 85\% on the hold-out test set which contains events equally distributed between the four classes of glitch, BBH, NSBH, and BNS. The high accuracy shows that the model effectively fits the data, enabling it to differentiate between all four classes. This is also apparent if we evaluate the model's performance by aggregating the multi-class predictions for 3 out of the 4 classes and considering it as a binary one-vs-all classifier for each combination: Glitch-vs-all, BBH-vs-all, NSBH-vs-all, and BNS-vs-all (i.e., comparing probability $p(class)$ to $p(not ~class)$ and using a threshold of 0.5 for classification). In all four cases the test accuracy is in the $\sim 90-95\%$ range, confirming that the model is generalizing equally well across classes (within small variations of a few percent).

The one-vs-all aggregation also allows us to compare the model's accuracy explicitly with previous versions of the model, which used a series of three one-vs-all classifiers trained separately: Glitch-vs-all, BBH-vs-all, and NS-vs-all. For each model version the test accuracy is evaluated on a 10\% hold-out set of events drawn from the same distribution as its training data. For the previous one-vs-all model versions, the classification threshold was set so that the test set false negative rate and false positive rate were equal (see Figure 4 in \cite{Abbott2022}. The comparisons in Figure \ref{fig:fig3_test_accuracy} show that when only the dataset is updated (as described in Section \ref{sec:data_updates}) while keeping the architecture the same as the original \gwsnm ~model \citep{Abbott2022}, the test accuracy across all three binary classifiers decreases on the order of $\sim 5-10\%$. This is likely due to the fact that in the updated dataset the SNR threshold of events is lower, and so includes more events that are harder to distinguish from background noise and thus may have fewer distinct features that the model can learn. Furthermore, the updated dataset for glitch events has a FAR cutoff of 2/day and so does not include some of the loudest and most obvious of glitch events, whereas there was no FAR threshold used in training \gwsnm. As discussed in Section~\ref{sec:data_updates}, this change is most apparent in the distribution of the Log BCI input value, with \gwsnm ~trained on glitches that generally have smaller Log BCI values as compared to astrophysical events, while \gwsnmm ~is trained on glitches and astrophysical events that have very similar distributions of the Log BCI values (see Figure~\ref{fig:fig2_inputs_distributions}(d)).

The updates made to the model architecture for \gwsnmm ~increase the test accuracy by a significant amount after the dataset updates have been implemented, such that each of the aggregated one-vs-all models have an accuracy that is either equal to (Glitch-vs-all) or higher than (BBH-vs-all, NSBH-vs-all, BNS-vs-all) the corresponding model with dataset updates only, demonstrating that the architecture changes contribute positively to the model's predictive power. While the percentage difference values are not large when comparing the test accuracy of \gwsnmm ~to \gwsnm, the accuracies are all above 90\%, showing that it is still in a high performance regime, with small trade-offs in optimizing astrophysical event predictions compared to glitches. Keeping in mind that the updated dataset is inherently harder to distinguish, while the updated model architecture is significantly simpler and provides more predictive information (distinguishes between NS events), \gwsnmm ~represents a significant update to the original model.

\subsection{Predictions for O3 Public Alerts}

There were 77 low-latency public alerts for candidate CBC events issued by the LVK in O3. Of these, two events, S190910h and S190930t, were single detector (Livingston) events. Since \gwsnmm ~is trained to predict on multi-detector events, we evaluate the model's performance on the 75 multi-detector O3 events only. The predictions are made using the low-latency localization maps that were generated by \texttt{BAYESTAR} and released in the O3 public alerts, as available on GraceDB \footnote{\url{https://gracedb.ligo.org/superevents/public/O3/}}. The prediction results from \gwsnmm ~are compared to the true classifications as determined in the full offline analysis of these events by the LVK and published in the final O3 catalogs GWTC-2.1 and GWTC-3 \citep{Abbott2023_gwtc3,Abbott2024_gwtc2.1} (hereafter collectively referred to as GWTC-3 for simplicity). For our analysis, only confident events determined with $p_{\mathrm{astro}} > 0.5$ and included in GWTC-3 are considered astrophysical events, with the rest classified as glitches. Of the astrophysical events, the merger type is determined from each of the best-fit component masses of the binary: if the mass is $< 3 ~M_{\odot}$ then the component is determined to be an NS; otherwise it is a BH. Thus of the 75 O3 public alerts analyzed, we determine from GWTC-3 that 32 are glitches (which includes 22 alerts that were subsequently retracted by the LVK), 40 are BBH, 2 are NSBH, and 1 is a BNS merger. The \gwsnmm ~classification is determined from the model's four output probabilities, selecting the class label with the highest predicted probability.

\begin{figure}
\includegraphics[width=\columnwidth]{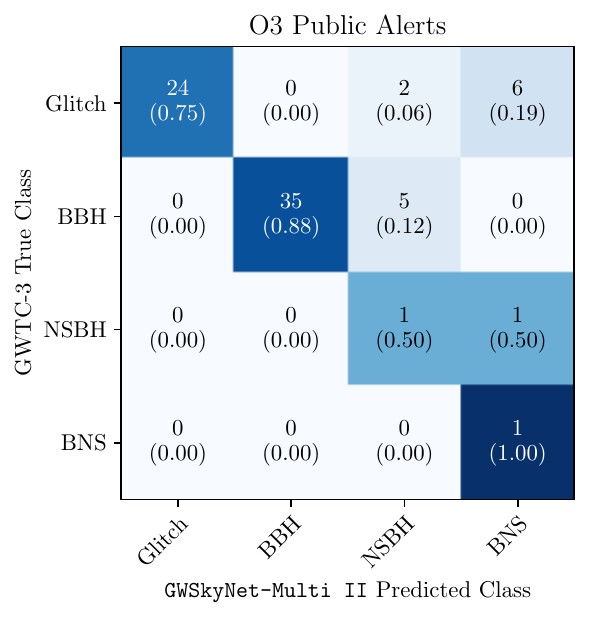}
\caption{Predicted versus true classification confusion matrix for \gwsnmm ~on O3 public alerts. \gwsnmm ~correctly predicts 61/75 (81\%) events, with most misclassifications being glitches identified as BNS (6 events), and BBH identified as NSBH (5 events). Crucially, the model does not produce false negative predictions for more promising events, indicating that for O3 alerts it is conservative in ``demoting'' an event that has a higher potential of having an electromagnetic counterpart to an event class with lower potential.
\label{fig:fig4_O3_confusion_matrix}}
\end{figure}

A comparison of the \gwsnmm ~predicted classification to the true GWTC-3 classification is shown in Figure \ref{fig:fig4_O3_confusion_matrix}. Overall, \gwsnmm ~correctly predicts 61/75 (81\%) of O3 events. If we only consider the model accuracy for determining glitch-vs-real, i.e., aggregate the astrophysical source classes, then it is able to correctly predict 67/75 (89\%) of O3 events.

We note that the off-diagonal values are all zero for the lower half of the confusion matrix, and the misclassifications are concentrated in the upper half of off-diagonal elements. As the classifications shown in Figure \ref{fig:fig4_O3_confusion_matrix} are ordered from lowest EM counterpart potential and follow-up priority (glitch) to highest (BNS), this indicates that for O3 events \gwsnmm ~does not produce false negative predictions for promising events, and tends to be conservative in making predictions for an event that would lower its potential for follow-up (i.e., false positives for high potential events are preferred over false negatives). For example, for true NSBH events, \gwsnmm ~only predicts them to be an NSBH or a higher priority BNS event. Conversely, if there is a public alert for a potential NSBH event, but it is actually a glitch, then \gwsnmm ~predicts it to be a glitch only when it is confident that it is not an NSBH or BNS. These results are currently only indicative of any underlying trends the model has learned, as the total number of NSBH and BNS events in O3 is too low for any robust conclusions. A full analysis of the final O4 results will help affirm (or deviate from) these trends.

For these same 75 events, we also determine the preliminary LVK classification at the time of the event from the \textit{p\_astro} probability information released in low-latency (and available on GraceDB), selecting the source type with the highest predicted probability as the LVK preliminary class. A comparison with the final LVK classification reveals that the preliminary classification correctly identified 51/75 (68\%) of events, an overall lower accuracy as compared to \gwsnmm ~(81\%). The vast majority of the preliminary misclassifications were glitches that were identified as astrophysical events. The preliminary LVK classifications in O3 included a fifth mutually exclusive source class, \textit{MassGap}, to identify events that might have merger components in the $3 - 5 ~M_{\odot}$ hypothetical ``mass gap'' range between NS and BH. This classification identified 11 events as most likely MassGap, of which the final classification revealed 4 to be glitches, 5 BBH, and 2 NSBH. As we do not target MassGap events in this work, and to simplify the comparison to the final classifications, we treat the 5 BBH and 2 NSBH events that were annotated as MassGap as being correctly identified when calculating the accuracy of the LVK preliminary classifications.

\subsubsection{Analysis of Misclassified Events}

A detailed comparison of O3 events misclassified by either \gwsnm ~or \gwsnmm ~is shown in Table \ref{table:O3_misclassified_predictions}. Both models have a comparable predictive accuracy, with \gwsnm ~correctly predicting 60/75 (80\%) events and \gwsnmm ~correctly predicting 61/75 (81\%) events. There are 9 events that were misclassified by \gwsnm ~but are correctly classified by \gwsnmm. This crucially includes the BNS merger event S190425z (discussed in detail in the next subsection). 6 events are misclassified by both models; these are predicted to be NS events by \gwsnm ~and BNS or NSBH events by \gwsnmm. And finally there are 8 events that were correctly classified by \gwsnm ~but are now misclassified by \gwsnmm. These are all glitch or BBH events that are misclassified as NSBH or BNS events, as well as one NSBH event that is misclassified as BNS (S200115j, discussed in detail in the next subsection). With the NS branch split into the BNS and NSBH branches, it becomes clear that the 5 BBH events that are being misclassified by \gwsnmm ~are all predicted to be NSBH. If we compare the mean distance input of these misclassifications, as shown in Figure \ref{fig:fig6_O3_classifications_vs_inputs} in Appendix~\ref{sec:appB}, we find that all 5 of these misclassified BBH events lie at the lower end of the BBH distribution.

\begin{table*}
\centering
\begin{tabular}{|l|c|c|c|c|c|c|c|}
\toprule
\toprule
\multicolumn{1}{|c|}{\textbf{Event ID}} & \multicolumn{5}{|c|}{\textbf{GWSkyNet-Multi II Prediction (\%)}} & \multicolumn{1}{|c|}{\textbf{GWSkyNet-Multi}}  & \multicolumn{1}{|c|}{\textbf{GWTC-3}} \\
\midrule
 & Glitch & BBH & NSBH & BNS & Classification & Classification & Classification \\
\midrule
\midrule
\textit{S190405ar} & $98 \pm 1$ & $0 \pm 0$ & $0 \pm 0$ & $2 \pm 1$ & Glitch & \textit{NS} & Glitch \\
\textit{S190425z} & $27 \pm 24$ & $1 \pm 2$ & $6 \pm 7$ & $66 \pm 22$ & BNS & \textit{Glitch} & BNS \\
\textit{S190630ag} & $2 \pm 3$ & $98 \pm 3$ & $0 \pm 0$ & $0 \pm 0$ & BBH & \textit{Glitch} & BBH \\
\textit{S191213g} & $93 \pm 8$ & $0 \pm 0$ & $0 \pm 0$ & $7 \pm 8$ & Glitch & \textit{NS} & Glitch \\
\textit{S191216ap} & $2 \pm 1$ & $94 \pm 3$ & $4 \pm 2$ & $0 \pm 0$ & BBH & \textit{NS} & BBH \\
\textit{S191220af} & $64 \pm 16$ & $0 \pm 0$ & $1 \pm 1$ & $35 \pm 15$ & Glitch & \textit{NS} & Glitch \\
\textit{S191225aq} & $70 \pm 27$ & $1 \pm 2$ & $2 \pm 6$ & $27 \pm 25$ & Glitch & \textit{NS} & Glitch \\
\textit{S200112r} & $6 \pm 7$ & $94 \pm 7$ & $0 \pm 0$ & $0 \pm 0$ & BBH & \textit{Glitch} & BBH \\
\textit{S200302c} & $1 \pm 1$ & $99 \pm 1$ & $0 \pm 0$ & $0 \pm 0$ & BBH & \textit{Glitch} & BBH \\
\midrule
\textbf{\textit{S190426c}} & $28 \pm 11$ & $0 \pm 0$ & $4 \pm 2$ & $68 \pm 11$ & \textbf{BNS} & \textit{NS} & Glitch \\
\textbf{\textit{S190503bf}} & $1 \pm 1$ & $1 \pm 1$ & $83 \pm 11$ & $15 \pm 10$ & \textbf{NSBH} & \textit{NS} & BBH \\
\textbf{\textit{S190816i}} & $6 \pm 4$ & $0 \pm 1$ & $43 \pm 11$ & $50 \pm 11$ & \textbf{BNS} & \textit{NS} & Glitch \\
\textbf{\textit{S190923y}} & $15 \pm 9$ & $0 \pm 0$ & $18 \pm 11$ & $67 \pm 12$ & \textbf{BNS} & \textit{NS} & Glitch \\
\textbf{\textit{S190924h}} & $0 \pm 0$ & $2 \pm 1$ & $97 \pm 2$ & $0 \pm 1$ & \textbf{NSBH} & \textit{NS} & BBH \\
\textbf{\textit{S200106au}} & $12 \pm 19$ & $0 \pm 1$ & $4 \pm 4$ & $84 \pm 18$ & \textbf{BNS} & \textit{NS} & Glitch \\
\midrule
\textbf{S190521g} & $0 \pm 0$ & $13 \pm 7$ & $87 \pm 7$ & $0 \pm 0$ & \textbf{NSBH} & BBH & BBH \\
\textbf{S190602aq} & $0 \pm 0$ & $28 \pm 9$ & $71 \pm 9$ & $0 \pm 0$ & \textbf{NSBH} & BBH & BBH \\
\textbf{S190829u} & $18 \pm 17$ & $1 \pm 2$ & $7 \pm 6$ & $74 \pm 16$ & \textbf{BNS} & Glitch & Glitch \\
\textbf{S190930s} & $1 \pm 0$ & $6 \pm 3$ & $94 \pm 3$ & $0 \pm 0$ & \textbf{NSBH} & BBH & BBH \\
\textbf{S191120aj} & $1 \pm 2$ & $0 \pm 0$ & $14 \pm 11$ & $85 \pm 11$ & \textbf{BNS} & Glitch & Glitch \\
\textbf{S200105ae} & $3 \pm 4$ & $1 \pm 2$ & $53 \pm 17$ & $44 \pm 18$ & \textbf{NSBH} & Glitch & Glitch \\
\textbf{S200108v} & $4 \pm 4$ & $8 \pm 9$ & $87 \pm 9$ & $1 \pm 3$ & \textbf{NSBH} & Glitch & Glitch \\
\textbf{S200115j} & $1 \pm 1$ & $0 \pm 0$ & $8 \pm 5$ & $90 \pm 6$ & \textbf{BNS} & NS & NSBH \\
\bottomrule
\end{tabular}
\caption{A comparison of O3 OPA events misclassified by \gwsnm ~(italicized), \gwsnmm ~(bolded), or both (bolded and italicized). The GWTC-3 classifications are obtained by considering only events with $p_{\mathrm{astro}} > 0.5$ as astrophysical (with the rest classified as glitches). The type of binary for the astrophysical events is determined from the best-fit component masses, assuming a maximum NS mass of $3 ~M_{\odot}$. The event S200105ae has $p_{\mathrm{astro}} = 0.36$ and so we consider it a glitch, but is reported in GWTC-3 as a \textit{marginal} NSBH candidate since it does pass the FAR threshold of $\mathrm{< 2/yr}$. Note in particular the predictions for event S190425z, which is misclassified by \gwsnm ~as a glitch, but is correctly classified as a BNS by \gwsnmm.}
\label{table:O3_misclassified_predictions}
\end{table*}

The most dominant misclassifications are of glitch events that are being predicted as NSBH or BNS events. This is a pattern that is seen both with \gwsnm ~and \gwsnmm. For the former, \cite{Raza2024} found that this arose because many of these glitch events had large Log BCI values and smaller sky localization areas, both of which the model had learned to associate with astrophysical events. However, when we attempt to compare the same inputs for the \gwsnmm ~misclassifications, we find no obvious trend (see Appendix~\ref{sec:appB}). This suggests that the model does not have a linear dependence on any one input, but has learned to distinguish classes from a non-linear combination of the inputs as codified in the updated neural network architecture. A deeper analysis of the model with machine learning interpretability tools could provide insights into its learned decision boundaries, and will be explored in future work.

The fact that \gwsnmm ~misclassifies some of the smaller distance BBH events as NSBH, and some of the glitch events as NSBH or BNS, is not ideal for EM follow-up campaigns. Targeting events that in reality have no (or low) potential for having an EM counterpart would lead to a waste of follow-up resources. However, since BNS and NSBH events are rare and have much higher scientific potential, a classifier that produces more false positives for NS events is a better outcome than one that produces false negatives (i.e, potentially misses high interest sources).

We also dis-aggregate the events and misclassifications according to the detector network type: 6/43 (14\%) of the 3-detector HLV events are misclassified, and 8/32 (25\%) of the 2-detector events are misclassified. Thus \gwsnmm ~misclassifies a larger percentage of 2-detector events as compared to 3-detector events. Within the 2-detector events, the fraction of misclassifications are: 5/19 (26\%) HL, 3/11 (27\%) LV, and 0/2 (0\%) HV, indicating that \gwsnmm ~has not learned any particular bias for 2-detector combinations. We contrast these to the misclassifications from \gwsnm: 5/43 (12\%) HLV, 2/19 (11\%) HL, 6/11 (55\%) LV, and 2/2 (100\%) HV, which clearly show a bias for misclassifying 2-detector events involving Virgo. This bias, identified in \cite{Raza2024}, is thus resolved in the updated model by including more glitch events in the training data that involve the Virgo detector.

\begin{table*}
\centering
\hspace*{-3.4cm}
\begin{tabular}{|l|c|c|c|c|c|c|c|c|}
\toprule
\toprule
\multicolumn{1}{|c|}{\textbf{Event ID}} & \multicolumn{4}{|c|}{\textbf{GWSkyNet-Multi II Probability (\%)}} & \multicolumn{4}{|c|}{\textbf{GWTC-3 Probability (\%)}} \\
\midrule
 & ~~Glitch~~ & ~~BBH~~ & ~~NSBH~~ & BNS & Glitch & BBH & NSBH & BNS \\
\midrule
\midrule
S190425z & $27 \pm 24$ & $1 \pm 2$ & $6 \pm 7$ & $66 \pm 22$ & 31 & 0 & 0 & 69 \\
S190814bv & $0 \pm 0$ & $18 \pm 9$ & $81 \pm 9$ & $0 \pm 1$ & 0 & 14 & 86 & 0 \\
S200115j & $1 \pm 1$ & $0 \pm 0$ & $8 \pm 5$ & $90 \pm 6$ & 0 & 0 & 100 & 0 \\
S200105ae & $3 \pm 4$ & $1 \pm 2$ & $53 \pm 17$ & $44 \pm 18$ & 64 & 0 & 36 & 0 \\
\bottomrule
\end{tabular}
\caption{A comparison of the classification probability scores of the three O3 OPA events that are included in the final GWTC-3 catalog as confident merger events involving NS, plus the marginal event S200105ae. \gwsnmm ~correctly classifies the first two events with the classification probabilities consistent with those in the GWTC-3 catalog. However, events S200115j and S200105ae are mis-classified (discussed in more detail in Section~\ref{sec:results_O3_NS}).}
\label{table:O3_NS_comparisons}
\end{table*}

\subsubsection{Analysis of NS Merger Events}\label{sec:results_O3_NS}

Three of the O3 alerts are included in the final GWTC-2.1 and GWTC-3 catalogs as confident astrophysical events that involve an NS: i) S190425z (GW190425) as a BNS, ii) S190814bv (GW190814) as an NSBH, and iii) S200115j (GW200115) as an NSBH. In addition, there is one marginal NSBH candidate S200105ae (GW200105). The event is considered marginal because it does not pass the threshold $p_{\mathrm{astro}} > 0.5$, but it does pass the threshold of $\mathrm{FAR < 2/yr}$. As these merger events (potentially) involve an NS, they are the most promising events of the O3 public alerts that could have an EM counterpart for follow-up observations. For these four events, we compare the corresponding GWTC-3 class probabilities (as determined by the detection pipeline with the highest SNR, see Table~XIII and Table~XV in \cite{Abbott2023_gwtc3}) and the predicted class probabilities by \gwsnmm, in Table~\ref{table:O3_NS_comparisons}. While only 4 events is not enough to draw conclusive statements, the comparisons serve to provide some insight into the model through illustrative examples.

For events S190425z and S190814bv, the probability scores that \gwsnmm ~predicts and the source probabilities determined by the detecting pipeline in the GWTC-3 final offline analysis are consistent. S190425z is classified by \gwsnmm ~as a BNS with $p_{\mathrm{BNS}} = 66 \pm 22\%$, compared to the GWTC-3 probability $p_{\mathrm{BNS}} = 69\%$ (detection pipelines used in GWTC-3 do not report uncertainties). While the uncertainty in the \gwsnmm ~prediction for this event is relatively high, the mean prediction probability is very close to the GWTC-3 value. For S190814bv the \gwsnmm ~classification is an NSBH with $p_{\mathrm{NSBH}} = 81 \pm 9\%$, compared to the GWTC-3 probability $p_{\mathrm{NSBH}} = 86\%$. The probability scores for the other three classes for this event also agree. The fact that \gwsnmm ~is able to give remarkably similar classification probabilities for these two events based on the preliminary low-latency analysis and localization data only, as compared to the final analysis of the fully calibrated strain data in GWTC-3, illustrates its predictive accuracy and utility as a classifier for making rapid follow-up decisions.

On the other hand, the \gwsnmm ~probability of NSBH for S200115j, $p_{\mathrm{NSBH}} = 8 \pm 5 \%$, is not consistent with the catalog value of $p_{\mathrm{NSBH}} = 100\%$. To get a sense of why \gwsnmm ~is mispredicting this event as a BNS we look at the masses of the components as determined in \cite{Abbott2023_gwtc3}: $m_1 = 5.9^{+2.0}_{-2.5} ~M_{\odot}$, $m_2 = 1.44^{+0.85}_{-0.28} ~M_{\odot}$, with chirp mass $\mathcal{M_C} = 2.43^{+0.05}_{-0.07} ~M_{\odot}$. The mass of the primary component is at the lower end of the stellar BH mass range, with a $29\%$ probability that $m_1 < 5 ~M_{\odot}$ and lies within the lower BH mass gap \citep{Abbott2023_gwtc3}. This implies that the BH in this merger is one of the lightest BHs detected in CBCs, and thus the event characteristics might be similar to some of the high-mass BNS events that \gwsnmm ~is trained on. The fact that the masses of the system lie close to the boundary of what the model has learned to distinguish between BNS and NSBH events is what leads \gwsnmm ~to mis-predict S200115j as a BNS event with $p_{\mathrm{BNS}} = 90 \pm 6 \%$.

For the marginal event S200105ae, \gwsnmm \space incorrectly predicts the event to be astrophysical with $p_{\mathrm{glitch}} = 3 \pm 4\%$, compared to the GWTC-3 probability of $p_{\mathrm{glitch}} = 64\%$. However, we note that the probabilities of the event being an NSBH merger are closer and marginally overlap within the associated uncertainty: $p_{\mathrm{NSBH}} = 53 \pm 17\%$, compared to the GWTC-3 $p_{\mathrm{NSBH}} = 36\%$. As more NSBH events are detected in future observations and we gain a better understanding of the rate of NSBH events, we will have a clearer picture of whether this event is truly a glitch or a weak NSBH signal that is difficult to distinguish from the noise.

\section{O4 Public Alerts} \label{sec:O4_predictions}

The LVK fourth observing run, O4, began in May 2023 and is expected to run until November 2025, sub-divided into three periods\footnote{\url{https://observing.docs.ligo.org/plan/}}. The first period, O4a, covers the time span from May 2023 to January 2024. During this time only the two LIGO detectors, Hanford and Livingston, were in science-observing mode and made GW detections. The LIGO detectors were joined by the Virgo detector in April 2024, marking the start of O4b, which ran until January 2025. The final period, O4c, began in late January 2025, and is expected to last until November 2025. The LVK has continued to release low-latency public alerts to the community throughout O4, dividing the alerts into two categories: i) significant candidate events, which have been detected with a false alarm rate FAR $\mathrm{< 2/yr}$, and ii) low significance events, which have FAR $\mathrm{< 2/day}$.

For the significant event alerts, the collaboration also provides follow-up alerts for the events, on a time-scale of hours to days, updating the event classification based on human vetting and a more refined analysis \citep{Chaudhary2024}. If the event is determined to arise from terrestrial sources, the alert is retracted and the event is classified as a glitch. Otherwise, a more detailed source parameter estimation is performed when possible, and an update alert is issued with improved estimates of the alert contents, including the latest predicted LVK classification probabilities. For low significance alerts the LVK does not provide follow-up analysis in low-latency. Final classifications for all candidate events will only become available once the collaboration releases its final O4 event catalogs.

\subsection{Predictions and analysis for O4 events}

In Table~\ref{table:O4_significant_predictions} we provide the \gwsnmm ~predictions for select significant event alerts that have been issued during O4a and O4b. A regularly updated list of all O4 event predictions, including low significance events, is made publicly available and maintained by the authors\footnote{\url{https://nayyer-raza.github.io/projects/GWSkyNet-Multi/}}. For events that were found by multiple search pipelines (superevents with multiple associated events), the predictions (and comparisons) shown are for the final preferred events as identified in low-latency by the LVK\footnote{\url{https://emfollow.docs.ligo.org/userguide/analysis/superevents.html}}. As done for the O3 analysis, the \gwsnmm ~classification is determined from the model's four output probabilities, selecting the class label with the highest predicted probability. Similarly, the LVK classification for each event is determined from the \textit{p\_astro} categorical probability, which is included in all CBC alert notices (except for retractions), selecting the source type with the highest estimated probability as the LVK predicted class. This is done for both the LVK preliminary alert classification and the updated classification.

There have been 195 multi-detector candidate CBC events that have associated \texttt{BAYESTAR} localization maps published in O4a and O4b (May 2023 - January 2025). Of these, there were 18 events that were subsequently retracted as glitch events, and 3 events that were not retracted but have an updated classification of being a glitch (including the event S240422ed \citep{LVK_GCN2024}). Three of the alerts that were not retracted have updated LVK probabilities that are consistent with being classified as NSBH. For the rest of the 171 events, the updated LVK classification is a BBH.

\setlength{\LTcapwidth}{\textwidth}
\begin{longtable*}{|l|c|c|c|c|c|c|c|c|c|c|c|}
\caption{Predictions for select significant O4 events} \\
\toprule
\toprule
\multicolumn{1}{|c|}{\textbf{Event ID}} & \multicolumn{5}{|c|}{\textbf{GWSkyNet-Multi II Prediction (\%)}} & \multicolumn{1}{|c|}{\textbf{LVK Prelim.}} & \multicolumn{5}{|c|}{\textbf{LVK Updated Prediction (\%)}} \\
\midrule
 & Glitch & BBH & NSBH & BNS & Class & Class & Glitch & BBH & NSBH & BNS & Class \\
\midrule
\midrule
\textbf{S230518h} & $0 \pm 0$ & $1 \pm 1$ & $28 \pm 8$ & $71 \pm 8$ & \textbf{BNS} & NSBH & 10 & 4 & 86 & 0 & NSBH \\
$\mathrm{S230524x^{\ddagger*}}$ & $100 \pm 0$ & $0 \pm 0$ & $0 \pm 0$ & $0 \pm 0$ & Glitch & $\mathrm{BNS^{\ddagger*}}$ & 100 & 0 & 0 & 0 & $\mathrm{Glitch^\dagger}$ \\
\textbf{S230622ba$\mathrm{^*}$} & $0 \pm 0$ & $2 \pm 2$ & $92 \pm 6$ & $6 \pm 6$ & \textbf{NSBH} & $\mathrm{BBH^*}$ & 100 & 0 & 0 & 0 & $\mathrm{Glitch^\dagger}$ \\
S230627c & $0 \pm 1$ & $37 \pm 21$ & $63 \pm 21$ & $0 \pm 0$ & NSBH & NSBH & 3 & 48 & 49 & 0 & NSBH \\
S230630am & $19 \pm 10$ & $81 \pm 10$ & $0 \pm 0$ & $0 \pm 0$ & BBH & BBH & 2 & 98 & 0 & 0 & BBH \\
S230630bq & $2 \pm 2$ & $74 \pm 8$ & $24 \pm 8$ & $0 \pm 0$ & BBH & BBH & 3 & 97 & 0 & 0 & BBH \\
S230706ah & $12 \pm 9$ & $87 \pm 9$ & $0 \pm 0$ & $0 \pm 0$ & BBH & BBH & 3 & 97 & 0 & 0 & BBH \\
S230708z & $19 \pm 7$ & $81 \pm 7$ & $0 \pm 0$ & $0 \pm 0$ & BBH & BBH & 5 & 95 & 0 & 0 & BBH \\
\textbf{S230708bi$\mathrm{^*}$} & $7 \pm 10$ & $77 \pm 18$ & $15 \pm 19$ & $0 \pm 0$ & \textbf{BBH} & $\mathrm{BBH^*}$ & 100 & 0 & 0 & 0 & $\mathrm{Glitch^\dagger}$ \\
\textbf{S230712a$\mathrm{^*}$} & $3 \pm 3$ & $31 \pm 21$ & $65 \pm 21$ & $0 \pm 0$ & \textbf{NSBH} & $\mathrm{BBH^*}$ & 100 & 0 & 0 & 0 & $\mathrm{Glitch^\dagger}$ \\
$\mathrm{S230715bw^*}$ & $81 \pm 17$ & $0 \pm 1$ & $17 \pm 16$ & $2 \pm 2$ & Glitch & $\mathrm{NSBH^*}$ & 100 & 0 & 0 & 0 & $\mathrm{Glitch^\dagger}$ \\
S230723ac & $1 \pm 1$ & $99 \pm 1$ & $1 \pm 1$ & $0 \pm 0$ & BBH & BBH & 13 & 87 & 0 & 0 & BBH \\
S230731an & $0 \pm 0$ & $96 \pm 2$ & $4 \pm 2$ & $0 \pm 0$ & BBH & BBH & 0 & 81 & 18 & 0 & BBH \\
S230807f & $14 \pm 6$ & $86 \pm 6$ & $0 \pm 0$ & $0 \pm 0$ & BBH & BBH & 5 & 95 & 0 & 0 & BBH \\
$\mathrm{S230810af^{\ddagger*}}$ & $97 \pm 13$ & $0 \pm 0$ & $0 \pm 0$ & $3 \pm 13$ & Glitch & $\mathrm{BNS^{\ddagger*}}$ & 100 & 0 & 0 & 0 & $\mathrm{Glitch^\dagger}$ \\
\textbf{S230830b$\mathrm{^*}$} & $3 \pm 4$ & $0 \pm 0$ & $9 \pm 5$ & $88 \pm 6$ & \textbf{BNS} & $\mathrm{NSBH^*}$ & 100 & 0 & 0 & 0 & $\mathrm{Glitch^\dagger}$ \\
$\mathrm{S230918aq^{\ddagger*}}$ & $100 \pm 0$ & $0 \pm 0$ & $0 \pm 0$ & $0 \pm 0$ & Glitch & $\mathrm{BNS^{\ddagger*}}$ & 100 & 0 & 0 & 0 & $\mathrm{Glitch^\dagger}$ \\
S230922q & $0 \pm 0$ & $79 \pm 8$ & $21 \pm 8$ & $0 \pm 0$ & BBH & BBH & 0 & 100 & 0 & 0 & BBH \\
$\mathrm{S231030av^{\ddagger*}}$ & $93 \pm 16$ & $0 \pm 0$ & $0 \pm 1$ & $7 \pm 16$ & Glitch & $\mathrm{BNS^{\ddagger*}}$ & 100 & 0 & 0 & 0 & $\mathrm{Glitch^\dagger}$ \\
\textbf{S231112ag$\mathrm{^*}$} & $20 \pm 25$ & $80 \pm 25$ & $0 \pm 0$ & $0 \pm 0$ & \textbf{BBH} & $\mathrm{BBH^*}$ & 100 & 0 & 0 & 0 & $\mathrm{Glitch^\dagger}$ \\
S231113bw & $0 \pm 0$ & $98 \pm 1$ & $2 \pm 1$ & $0 \pm 0$ & BBH & BBH & 4 & 96 & 0 & 0 & BBH \\
S231118an & $0 \pm 0$ & $99 \pm 1$ & $1 \pm 1$ & $0 \pm 0$ & BBH & BBH & 24 & 74 & 1 & 0 & BBH \\
S231119u & $19 \pm 8$ & $81 \pm 8$ & $0 \pm 0$ & $0 \pm 0$ & BBH & BBH & 5 & 95 & 0 & 0 & BBH \\
\midrule
$\mathrm{S240413p^*}$ & $0 \pm 1$ & $95 \pm 4$ & $5 \pm 4$ & $0 \pm 0$ & BBH & $\mathrm{NSBH^*}$ & 2 & 98 & 0 & 0 & BBH \\
$\mathrm{S240420aw^*}$ & $53 \pm 29$ & $47 \pm 29$ & $0 \pm 0$ & $0 \pm 0$ & Glitch & $\mathrm{BBH^*}$ & 100 & 0 & 0 & 0 & $\mathrm{Glitch^\dagger}$ \\
$\textbf{S240421ar}$ & $42 \pm 11$ & $58 \pm 11$ & $0 \pm 0$ & $0 \pm 0$ & \textbf{BBH} & Glitch & 59 & 41 & 0 & 0 & Glitch \\
$\mathrm{S240422ed^*}$ & $58 \pm 17$ & $0 \pm 0$ & $1 \pm 1$ & $41 \pm 16$ & Glitch & $\mathrm{NSBH^*}$ & 93 & 0 & 2 & 5 & Glitch \\
\textbf{S240423br$\mathrm{^*}$} & $1 \pm 2$ & $99 \pm 2$ & $1 \pm 1$ & $0 \pm 0$ & \textbf{BBH} & $\mathrm{BBH^*}$ & 100 & 0 & 0 & 0 & $\mathrm{Glitch^\dagger}$ \\
\textbf{S240426s} & $2 \pm 2$ & $0 \pm 0$ & $33 \pm 13$ & $65 \pm 13$ & \textbf{BNS} & BBH & 2 & 98 & 0 & 0 & BBH \\
S240426dl & $73 \pm 12$ & $27 \pm 12$ & $0 \pm 0$ & $0 \pm 0$ & Glitch & Glitch & 70 & 30 & 0 & 0 & Glitch \\
$\mathrm{S240429an^{\ddagger}}$ & $85 \pm 26$ & $1 \pm 2$ & $2 \pm 6$ & $12 \pm 23$ & Glitch & $\mathrm{Glitch^{\ddagger}}$ & 100 & 0 & 0 & 0 & $\mathrm{Glitch^\dagger}$ \\
S240601aj & $4 \pm 3$ & $96 \pm 3$ & $0 \pm 0$ & $0 \pm 0$ & BBH & BBH & 49 & 51 & 0 & 0 & BBH \\
S240618ah & $13 \pm 6$ & $87 \pm 6$ & $0 \pm 0$ & $0 \pm 0$ & BBH & BBH & 4 & 96 & 0 & 0 & BBH \\
\textbf{S240621em} & $58 \pm 14$ & $42 \pm 14$ & $0 \pm 0$ & $0 \pm 0$ & \textbf{Glitch} & BBH & 4 & 96 & 0 & 0 & BBH \\
$\mathrm{S240623dg^{\ddagger}}$ & $70 \pm 37$ & $0 \pm 0$ & $1 \pm 2$ & $29 \pm 36$ & Glitch & $\mathrm{Glitch^{\ddagger}}$ & 100 & 0 & 0 & 0 & $\mathrm{Glitch^\dagger}$ \\
S240627by & $15 \pm 9$ & $79 \pm 9$ & $6 \pm 3$ & $0 \pm 0$ & BBH & BBH & 1 & 99 & 0 & 0 & BBH \\
S240825ar & $2 \pm 2$ & $85 \pm 5$ & $12 \pm 4$ & $0 \pm 0$ & BBH & BBH & 1 & 97 & 3 & 0 & BBH \\
S240830gn & $0 \pm 0$ & $94 \pm 3$ & $6 \pm 3$ & $0 \pm 0$ & BBH & BBH & 0 & 89 & 11 & 0 & BBH \\
S240910ci & $0 \pm 0$ & $99 \pm 1$ & $1 \pm 1$ & $0 \pm 0$ & BBH & BBH & 0 & 69 & 31 & 0 & BBH \\
S240915b & $0 \pm 0$ & $99 \pm 1$ & $0 \pm 0$ & $0 \pm 0$ & BBH & BBH & 0 & 86 & 14 & 0 & BBH \\
S240916ar & $31 \pm 13$ & $64 \pm 12$ & $5 \pm 3$ & $0 \pm 0$ & BBH & BBH & 1 & 99 & 0 & 0 & BBH \\
S240917cb & $37 \pm 14$ & $63 \pm 14$ & $0 \pm 0$ & $0 \pm 0$ & BBH & BBH & 4 & 96 & 0 & 0 & BBH \\
\textbf{S240925n} & $0 \pm 1$ & $49 \pm 15$ & $51 \pm 15$ & $0 \pm 0$ & \textbf{NSBH} & BBH & 0 & 100 & 0 & 0 & BBH \\
$\mathrm{S241005bo^{\ddagger}}$ & $99 \pm 4$ & $0 \pm 0$ & $0 \pm 0$ & $1 \pm 4$ & Glitch & $\mathrm{Glitch^{\ddagger}}$ & 100 & 0 & 0 & 0 & $\mathrm{Glitch^\dagger}$ \\
S241011k & $1 \pm 1$ & $89 \pm 7$ & $10 \pm 7$ & $0 \pm 0$ & BBH & BBH & 0 & 100 & 0 & 0 & BBH \\
\textbf{S241102br} & $0 \pm 1$ & $43 \pm 18$ & $57 \pm 18$ & $0 \pm 0$ & \textbf{NSBH} & BBH & 0 & 99 & 1 & 0 & BBH \\
S241104a & $83 \pm 9$ & $5 \pm 4$ & $13 \pm 6$ & $0 \pm 0$ & Glitch & Glitch & 100 & 0 & 0 & 0 & $\mathrm{Glitch^\dagger}$ \\
S241109bn & $2 \pm 1$ & $33 \pm 11$ & $65 \pm 11$ & $0 \pm 0$ & NSBH & NSBH & 0 & 28 & 72 & 0 & NSBH \\
\textbf{S241110br} & $4 \pm 2$ & $34 \pm 12$ & $62 \pm 13$ & $0 \pm 0$ & \textbf{NSBH} & BBH & 0 & 100 & 0 & 0 & BBH \\
S241114bi & $2 \pm 4$ & $81 \pm 8$ & $16 \pm 9$ & $0 \pm 0$ & BBH & BBH & 0 & 91 & 9 & 0 & BBH \\
$\mathrm{S241126dm^*}$ & $75 \pm 37$ & $7 \pm 13$ & $16 \pm 30$ & $2 \pm 7$ & Glitch & $\mathrm{BBH^*}$ & 100 & 0 & 0 & 0 & $\mathrm{Glitch^\dagger}$ \\
S241130be & $3 \pm 3$ & $82 \pm 6$ & $15 \pm 5$ & $0 \pm 0$ & BBH & BBH & 0 & 100 & 0 & 0 & BBH \\
S241201ac & $24 \pm 11$ & $76 \pm 11$ & $0 \pm 0$ & $0 \pm 0$ & BBH & BBH & 3 & 97 & 0 & 0 & BBH \\
S241210cw & $23 \pm 9$ & $77 \pm 9$ & $0 \pm 0$ & $0 \pm 0$ & BBH & BBH & 0 & 100 & 0 & 0 & BBH \\
S250101k & $1 \pm 1$ & $99 \pm 1$ & $0 \pm 0$ & $0 \pm 0$ & BBH & BBH & 4 & 88 & 8 & 0 & BBH \\
S250108ha & $92 \pm 4$ & $0 \pm 0$ & $1 \pm 1$ & $6 \pm 3$ & Glitch & Glitch & 100 & 0 & 0 & 0 & $\mathrm{Glitch^\dagger}$ \\
S250118az & $2 \pm 2$ & $88 \pm 4$ & $9 \pm 3$ & $0 \pm 0$ & BBH & BBH & 1 & 99 & 0 & 0 & BBH \\
\bottomrule
\caption{\gwsnmm ~predictions for select significant O4 events. Of the 195 multi-detector CBC candidate event alerts with \texttt{BAYESTAR} maps published during O4a and O4b (May 2023 - January 2025), we show all events except those for which both the LVK and \gwsnmm ~predicted probabilities of BBH are $> 90\%$ (i.e., we omit the 138 events that are confidently predicted as BBH by both classifications). The remaining 57 events are listed here. Also shown for each event are: i) the LVK preliminary alert classification (highest probability class), and ii) the LVK updated classification (with predicted probabilities) based on the latest alert issued for that event. Events for which the predicted \gwsnmm ~class and LVK updated class do not match are highlighted in bold (13/195). Events for which the LVK preliminary class and LVK updated class do not match are indicated with an asterisk ($*$) (15/195). If the LVK updated class is a glitch due to the event alert being retracted by the LVK, the prediction class is annotated with a dagger ($\dagger$) (18 events). If the event is an early warning alert (with the LVK preliminary class determined from the early warning alert) it is annotated with a double-dagger ($\ddagger$) (7 events). The horizontal line in the middle of the table indicates the boundary between the first two parts of the fourth observing run, O4a and O4b. Note in particular the event S240422ed, a high interest potential NSBH alert, which \gwsnmm ~predicted to likely be a glitch with a probability of 58\%, and the LVK updated the classification after further analysis to be consistent with a glitch as well.}
\label{table:O4_significant_predictions}
\end{longtable*}

\begin{figure}
\includegraphics[width=\columnwidth]{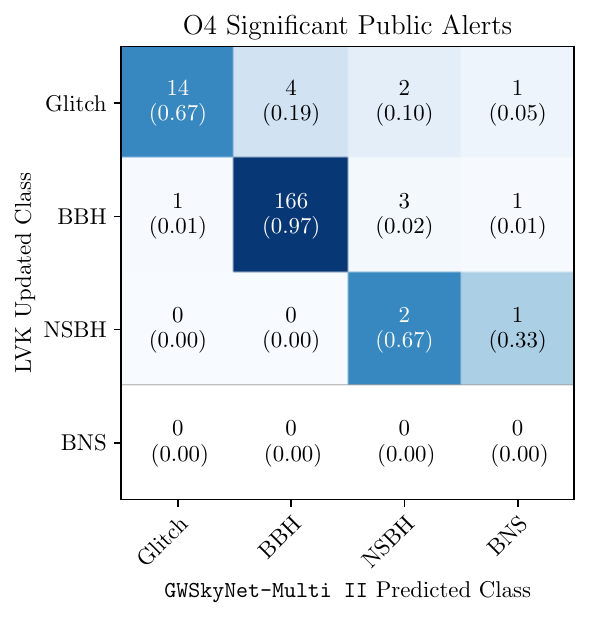}
\caption{Confusion matrix showing \gwsnmm ~predictions for all 195 multi-detector significant CBC events in O4a and O4b (May 2023 - January 2025), compared to the LVK updated (latest available) classification. This includes 18 events that have been retracted by the LVK as glitches, and 3 events that were not retracted but have an updated classification of being a glitch (see Table~\ref{table:O4_significant_predictions}). The \gwsnmm ~predicted class is consistent with the LVK updated class for 182/195 (93\%) events.
\label{fig:fig5_O4_confusion_matrix}}
\end{figure}

An analysis of the performance of \gwsnmm \ predictions for O4 events can be done by comparing the predictions to the latest LVK updated predictions. This is a preliminary analysis because the true classification of events can only be known after the LVK O4 catalog results become available. There is the possibility that some events are classified differently in the final catalogs as compared to the LVK updated classifications presented here. However, comparing to the latest LVK classification gives us a reasonable comparison metric and a check for our model's prediction validity. In Figure \ref{fig:fig5_O4_confusion_matrix} we provide a comparison of the \gwsnmm ~predicted class (highest probability class) to the LVK updated prediction class (highest probability class) for events in O4a and O4b. The \gwsnmm ~class is consistent with the LVK updated class for 182/195 (93\%) events. The vast majority (97\%) of candidate events identified as BBH in the LVK updated classification are also predicted to be BBH by \gwsnmm. Of the 3 events identified as NSBH in the LVK updated class, the model predicts 2 to be NSBH and 1 a BNS merger. For the 21 events that have an LVK updated class of being a glitch, while \gwsnmm ~predicts most to be glitches as well, a significant fraction (33\%) are classified as astrophysical events instead.

We note that of these 21 events with an LVK updated classification of glitch, 7 events are from early warning (pre-merger) alerts \citep{Magee2021} (these events are annotated with the $\ddagger$ symbol in Table~\ref{table:O4_significant_predictions}). All 7 of these candidate events were later retracted by the LVK, and \gwsnmm ~classifies them as a glitch as well. However, since \gwsnmm ~is not explicitly trained on early warning events, we refrain from making any robust claims on its performance for these types of events; this would require extensive testing on a suite of early warning alerts which also includes positive (non-glitch) examples, which we will explore in future work.

While the overall performance of the model in this initial analysis is promising --- identifying 93\% of events consistently with the LVK updated classification --- we leave a more detailed analysis of the accuracy of the predictions to future work, once the final LVK O4 catalogs are available and there is more complete information about the true nature of the event sources. The comparison to final classifications will help further illuminate the strengths and weaknesses of \gwsnmm.

\subsection{Using \gwsnmm ~as a follow-up tool}

When a preliminary alert for a candidate CBC event is issued by the LVK, it includes information about the LVK \textit{p\_astro} classification. For the significant O4a and O4b events, the LVK preliminary classifications match the updated classifications for 180/195 (92\%) events, and so the \gwsnmm ~predictions and the LVK preliminary predictions have an overall similar level of consistency with the LVK updated predictions. Compared to each other, the \gwsnmm ~classifications match the LVK preliminary classifications for 176/195 (90\%) events. There are 6 events for which both the \gwsnmm ~class and the LVK preliminary class are inconsistent with the LVK updated class; all 6 events have an LVK updated class of glitch (retracted) but are predicted to be astrophysical by both \gwsnmm ~and LVK preliminary (see events highlighted with bold and asterisks in Table~\ref{table:O4_significant_predictions}).

Events in O4 that pass a certain SNR threshold are also annotated by the \gwsn ~model, which has been integrated into the LVK low-latency analysis infrastructure, and the results made available on GraceDB \citep{Chan2024}. \gwsn ~is a binary classifier targeted towards and optimized for glitch-vs-real classifications, and has different training data, architecture, inputs, and reported outputs as compared to \gwsnmm. The model makes predictions for only those events that pass an SNR threshold of $\rho_{det} > 4.5$ in at least two detectors and a network SNR $\rho_{net} > 7$. Of the 195 significant CBC candidate alerts in O4a and O4b that are classified by \gwsnmm, 36 events do not pass this SNR threshold and are thus not annotated by \gwsn. For the remaining 159 events, \gwsn ~predicts 11 to be glitches and 148 to be astrophysical, assuming a classification threshold score of 0.75 (which balances an 80\% noise rejection rate with a 93\% rate of capturing astrophysical signals \citep{Chan2024}). For these same events \gwsnmm ~predicts 5 to be glitches and 154 to be astrophysical, the classification matching 3/11 \gwsn ~predicted glitches and 146/148 \gwsn ~predicted astrophysical events, for an overall consistency of 149/159 (94\%) events (where we've combined the BBH, NSBH, and BNS classifications into a single `astrophysical' class for comparison).

The comparisons of \gwsnmm ~to the LVK preliminary (\textit{p\_astro}) classifications and the \gwsn ~classifications highlight two key points: i) the vast majority ($>90\%$) of the predicted classifications from these different pipelines agree with each other, building confidence in their prediction validity, and ii) in cases where the predictions do not agree, \gwsnmm ~adds unique capabilities to aid astronomers in selecting targets for follow-up observations.

We select two events to highlight as examples of the model's capabilities: (1) The event S240422ed was a high interest candidate with a preliminary LVK classification of NSBH. \gwsnmm ~(and \gwsn) classified the event as a likely glitch, indicating that electromagnetic follow-up was not warranted. After further analysis, the LVK updated this candidate event's classification to glitch as well. (2) The event S240413p had a LVK preliminary classification of NSBH and a \gwsn ~classification of astrophysical. It was annotated as a BBH by \gwsnmm, and the LVK subsequently updated the classification to BBH as well. The \gwsnmm ~individual breakdown of classes added valuable information by distinguishing candidate BBH events from NSBH/BNS, helping to filter them from follow-up observations. Furthermore, there are candidate events that \gwsn ~does not annotate because they do not pass its SNR threshold (36/195 events in O4a + O4b). For these events \gwsnmm ~still provides a prediction (since it has been trained on lower SNR events), which can be considered alongside the LVK preliminary classifications for making rapid follow-up decisions.

The implementation of \gwsnmm ~in a low-latency scenario for follow-up observations requires that running the model not add significant delay once the LVK public alert is received. The median latency between the candidate merger time and the LVK preliminary alert being issued was 29.7 seconds for events in O4a \citep{Chaudhary2024}. We calculate the end-to-end time it takes to run \gwsnmm ~for each event in O4a and O4b, i.e, from when the LVK public alert is received to when the prediction classification is output to the user. The total time encompasses three main steps: i) downloading the BAYESTAR localization FITS file for the event from the GraceDB servers, ii) processing the data in the FITS file to extract and calculate the relevant model inputs, and iii) running the inputs through the model to make predictions (inference). We find a median end-to-end latency of 10 seconds when running on an Apple M1 Pro processor (released in 2021) and 18 seconds on an AMD Opteron 6282SE processor (released in 2011).

The latency timing shows that when run on modern machines, \gwsnmm ~can provide predictions for events without adding any significant delay to follow-up observations. As the model is both light-weight and fast, it is ideally suited for helping to make rapid decisions, and can easily be implemented by users to select for events in their own follow-up observation pipelines (for example, as a custom filter in the ANTARES alert broker \citep{Matheson2021} for the Vera C. Rubin Observatory \citep{Ivezic2019}). For users who wish to automate their follow-up observations for minimal latency, we recommend that they run \gwsnmm ~locally, using the model and scripts released in the associated repository \citep{Raza2025}. For users who can afford a higher latency and do not want to implement the model locally, but still want to know the model's predictions before making follow-up decisions, the authors maintain a webpage\footnote{\url{https://nayyer-raza.github.io/projects/GWSkyNet-Multi/}} with a table listing all significant O4 event predictions, which is typically updated $\sim 1$ minute after a LVK public alert is issued. The predictions table can be viewed directly on the webpage or downloaded as a Comma-Separated Values (CSV) file.

\section{Summary and Conclusion} \label{sec:conclusion}

In this work we have introduced \gwsnmm, a real-time machine learning classifier for gravitational-wave events that offers a significant update to the original model for predictions in O4 and beyond.

Motivated by the findings in \cite{Raza2024}, we update the training dataset to include examples of glitch events from the LVK O3 run, and simulate astrophysical events from updated population models. The detector network input is modified to more accurately mirror the information available in the public alerts, which allows us to include examples of glitches that were observed in the 3 detector HLV configuration. The SNR threshold for including events in the data is also reduced, to better capture low significance event alerts that are published by the LVK in O4. This provides a less biased and more representative sample of LVK events to train the model.

The model architecture is significantly simplified while making the predictions more informative. The three one-vs-all models in \gwsnm ~(glitch, BBH, NS) are replaced by a single multi-class classifier that gives normalized probability scores for all four CBC candidate event categories released by the LVK: glitch, BBH, NSBH, and BNS. Thus the model is now able to distinguish between NSBH and BNS events. The 2D sky map and 3D volume projection image inputs are replaced by the 90\% confidence interval values for the sky area and volume, respectively. This removes the convolutional branch of the model and reduces the inputs to numerical (tabular) data only that are more physically intuitive and interpretable to the user. The maximum distance is replaced by the distance uncertainty estimate, while the log BSN input is re-normalized to a maximum value of 100.

The architecture is simplified such that it has only 20 neurons in total: two fully connected hidden layers of 8 neurons each followed by an output layer of 4 neurons. This represents a factor of 60 reduction in the number of trainable parameters, as compared to \gwsnm. By taking an ensemble of predictions, the model also provides uncertainties associated with each classification probability, giving more information to the end user for how confident the model predictions are. The simplicity of the architecture allows for a significantly shorter training time and motivates the application of a wider variety of explainability tools in future studies of the model.

The updated model is evaluated on the hold-out test set of events and compared with the previous model. To perform a one-to-one comparison with \gwsnm \ the model predictions are evaluated in one-vs-all mode, and the accuracy is calculated. While the accuracy differences for the models are small when comparing \gwsnm ~to \gwsnmm, the accuracies are all above 90\%, indicating that the model remains in a high performance regime. Coupled with the fact that glitches in the updated dataset are harder to differentiate from the astrophysical events, and the model architecture is significantly simpler while giving more predictive information (splits the NS class and distinguishes between NSBH and BNS), the \gwsnmm ~model is a significant update to the original model.

A comparison of the \gwsnmm ~classifications to the true GWTC-3 classifications shows that the model correctly predicts 61/75 (81\%) of O3 events. \gwsnmm ~over-predicts the number of NSBH and BNS events, which are high potential targets for EM follow-up. On the other hand, it does not misclassify any real astrophysical events as glitches, nor any NSBH or BNS events as BBH. This means that the model is conservative in ``downgrading'' the classification of an event to a less promising class for follow-up. While the overall O3 accuracy is the same as for \gwsnm, the type of events that are misclassified are different. In particular the BNS event S190425z is correctly predicted by the new model, while it was classified as a glitch before. Analysis of the misclassifications shows that \gwsnmm ~does not have the same biases that are present in the original model, and in particular does not disproportionately struggle with events involving the Virgo detector. An analysis of the model dependence on the inputs in the context of misclassifications reveals no simple linear relationship or bias, suggesting that the model has learned non-linear feature representations through the combination of the inputs to distinguish classes.

With the updated model, we also provide the \gwsnmm ~predictions for select significant event alerts that have been issued during O4a and O4b (May 2023 to January 2025). A list of all O4 event predictions, including for the latest event alerts, is maintained by the authors and publicly available\footnote{\url{https://nayyer-raza.github.io/projects/GWSkyNet-Multi/}}. When compared to the latest updated LVK classifications, for an initial analysis of the model's performance, we find that the predicted \gwsnmm ~class is consistent with the LVK updated class for 182/195 (93\%) of these alerts, which is similar to the classification consistency of the LVK preliminary predictions (92\%). The \gwsnmm ~class also matches the \gwsn ~classification for 94\% of events predicted. While a complete analysis of the model's accuracy and robustness will be possible after the final LVK O4 catalog results becomes available, the consistency of \gwsnmm ~predictions in these initial comparisons is promising and illustrates the utility of the model. We recommend that the \gwsnmm ~predictions be considered together with the LVK preliminary and \gwsn ~predictions available on GraceDB at the time of the event, to maximize follow-up observation strategies and reduce misallocation of limited telescope time and resources.

Analysis of the \gwsnmm ~predictions on all the varied data sets across observing runs shows that the updated model provides more robust and informative predictions for use by the community to make follow-up observation decisions. Combined with the fact the the model is light-weight and fast, users can implement it in their own follow-up pipelines with minimal overhead and latency. A repository containing the updated model, together with scripts and instructions on how to use the model, is publicly available \citep{Raza2025}.

In future work we plan to study the model with machine learning explainability methods to shed further light on the model's inner workings and its learned decision boundaries. The promising performance of \gwsnmm ~on O4 early warning alerts, despite not explicitly being trained on such events, suggests the model's capabilities could be expanded in future observing runs to robustly classify these early warning merger alerts. Motivated by the updated model's ability to distinguish between the three astrophysical source types of BNS, NSBH, and BBH, we will also explore extending the classification scheme to label merger events with components in the  hypothetical ``mass gap'' between the heaviest known NS and lightest BH.

\vspace{9cm}

The authors wish to acknowledge and highlight the contributions of Miriam Cabero, Thomas Abbott, Eitan Buffaz, and Nicholas Vieira in developing the original version of the model, \gwsnm. The authors also wish to thank members of the ML-ESTEEM collaboration, in particular Fr\'{e}d\'{e}ric Beaupr\'{e}, Ren\'{e}e Hlo\v{z}eck, Flavie Lavoie-Cardinal, and Niko Lecoeuche, for their insights and discussions that helped guide this work.

The authors acknowledge support for this project from the Canadian Tri-Agency New Frontiers in Research Fund - Exploration program, and from the Canadian Institute for Advanced Research (CIFAR), in particular the CIFAR Catalyst program in supporting the ML-ESTEEM collaboration. N.R. is supported by a Walter C. Sumner Memorial Fellowship and acknowledges funding support from the Trottier Space Institute at McGill. D.H. and J.M. acknowledge support from the Natural Sciences and Engineering Research Council of Canada (NSERC) Discovery Grant program and the Canada Research Chairs (CRC) program. A.M. acknowledges support from the NSF (1640818, AST-1815034). A.M. and J.M. also acknowledge support from IUSSTF (JC-001/2017). This material is based upon work supported by NSF’s LIGO Laboratory, which is a major facility fully funded by the National Science Foundation.


\appendix

\section{Re-weighting NSBH and BNS events}\label{sec:appA}

The training dataset has equal numbers of samples for each of the four classes, but in reality their occurrence rates are quite different. In particular, NSBH and BNS merger events are rare compared to BBH mergers and glitches: of the 90 significant CBC events reported by the LVK between O1 and O3, 83 are confidently classified as BBH events ($M_1, M_2 > 3~M_{\odot}$) \citep{Abbott2019_gwtc1,Abbott2024_gwtc2.1,Abbott2023_gwtc3}. Similarly, of the 75 CBC candidate public alerts issued in O3, there were: 32 Glitch, 40 BBH, 2 NSBH, and 1 BNS. Providing a class-balanced dataset during training helps the model to learn the differences between the classes, but for our model this also leads to over-predicting the number of BNS and NSBH events as compared to the real number of O3 events.

We attempt to correct for this and find that down-weighting the NSBH events during training by a factor of 5 and the BNS events by a factor of 10, as compared to the BBH and glitch events, helps prevent the model from over-predicting the occurrence of these events. This has a marginal negative impact on the model's test set performance (which is still class balanced), but a significant positive effect on its performance on O3 public alerts, as seen in Table~\ref{table:reweighting_comparisons}, which shows experiments performed for a range of different down-weighting values. A down-weighting of BNS by a factor of 10 and NSBH by a factor of 5 strikes the right balance: 3\% loss in test accuracy, for a 16\% gain in O3 accuracy. The drop in performance between the test and O3 datasets is not completely surprising, since these follow different distributions, and the gap is indicative of the differences in the training dataset (which includes simulated events) and the real merger population. We find that the down-weighting helps the model learn different decision boundaries for input features, which generalize better to actual alerts in O3. Of course these preliminary results are optimistic (over-fitted to O3) since they rely on knowing the exact distribution of classes in O3. Generalization will be evaluated on the O4 dataset once the final catalogs are released by the LVK.

\begin{table}
\vspace*{0.2cm}
\centering
\begin{tabular}{ |c|c|c|c| }
\toprule
\toprule
\multicolumn{2}{|c|}{\textbf{Down-weighting Factor}} & \multicolumn{2}{|c|}{\textbf{Model Accuracy (\%)}} \\
\midrule
NSBH & BNS & Test Set & O3 Events \\
\midrule
\midrule
1 & 1 & $87.8 \pm 1.5$ & 65.3 \\
2 & 2 & $87.9 \pm 1.3$ & 69.3 \\
2 & 4 & $86.9 \pm 1.4$ & 69.3 \\
5 & 5 & $85.9 \pm 1.4$ & 76.0 \\
5 & 10 & $85.1 \pm 1.3$ & 81.3 \\
10 & 10 & $83.5 \pm 1.3$ & 81.3 \\
10 & 20 & $83.0 \pm 1.4$ & 81.3 \\
20 & 20 & $80.6 \pm 1.9$ & 81.3 \\
20 & 40 & $80.5 \pm 1.9$ & 82.7 \\
\bottomrule
\end{tabular}
\caption{Experiments in down-weighting the NSBH and BNS events in the model training (as compared to the BBH and glitch events), and their impact on the model's test set accuracy and O3 events accuracy. The drop of performance between the test and O3 datasets reflects how well the features learned by the model in the training generalize to the distribution of events in O3. A down-weighting of BNS by a factor of 10 and NSBH by a factor of 5 strikes the optimal balance: a marginal 3\% loss in test accuracy, for a 16\% gain in O3 accuracy.}
\label{table:reweighting_comparisons}
\end{table}

Instead of down-weighting, we could also have a non class-balanced training set, reducing the number of NSBH and BNS events (e.g., training on 150 BNS events instead of 1500). The disadvantage of this approach is that we lose variety in the type of BNS and NSBH events that the model can learn.

\section{O3 classifications analysis based on input values}\label{sec:appB}

\begin{figure*}
\includegraphics[width=\textwidth]{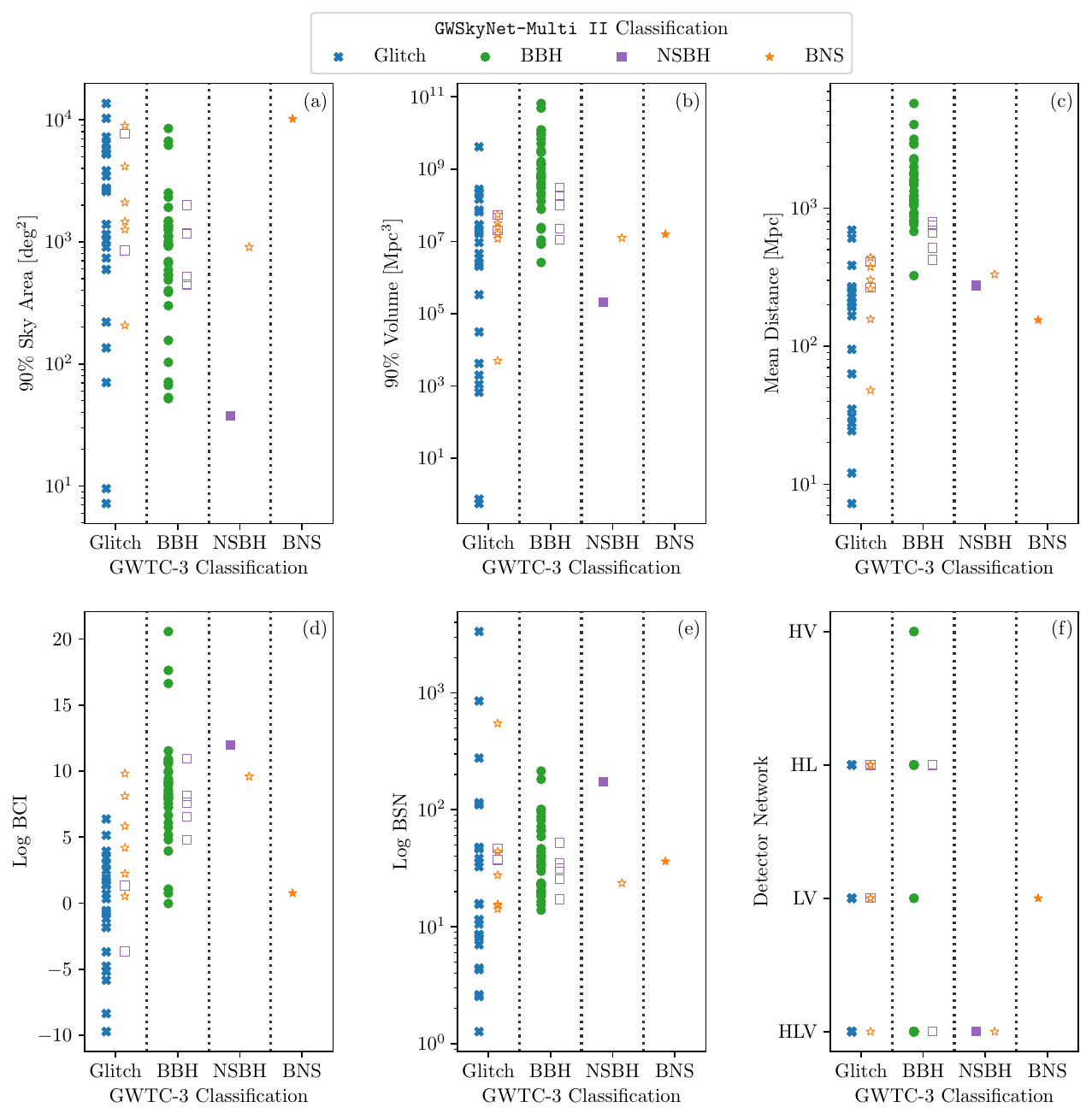}
\caption{Classifications of the 75 multi-detector candidate CBC events in O3 for which a public alert was issued, distributed vertically by the input parameters: (a) sky map localization area, (b) 3D volume localization, (c) estimated mean distance, (d) Bayes factor for coherence vs. incoherence, (e) Bayes factor for signal vs. noise, and (f) observing detector network. In each panel the events are horizontally divided according to their true classification from GWTC-3: 32 glitch events in the left column, 40 BBH events in the center-left column, 2 NSBH events in the center-right column, and 1 BNS event in the right column. The marker for each event is labeled according to its predicted classification from \gwsnmm: glitches in blue crosses, BBH in green circles, NSBH in purple squares, and BNS in orange stars. Events that are correctly classified by \gwsnmm ~have filled markers (e.g., the filled blue crosses for glitches in the glitch column), while misclassified events have open markers that are slightly offset to the right (e.g., the unfilled orange stars for BNS in the glitch column). The misclassified events are compared to the correct classifications to find regions in the input values where the misclassifications might stand out. Aside from the 5 BBH events that are misclassified as NSBH having mean distances at the lower end of the distribution, there are no other straightforward boundary regions identified in this analysis.
\label{fig:fig6_O3_classifications_vs_inputs}}
\end{figure*}

We provide a comparison of the O3 events correctly classified and misclassified by \gwsnmm ~in the context of the model input values in Figure \ref{fig:fig6_O3_classifications_vs_inputs}. Aside from the 5 BBH events that are misclassified as NSBH having mean distances at the lower end of the distribution, there are no other straightforward boundary regions identified in this analysis. This suggests that the model does not have a linear dependence on any one input, but distinguishes classes from a non-linear combination of the inputs codified in the neural network.

\bibliography{references}{}
\bibliographystyle{aasjournal}

\end{document}